\documentclass[useamsfonts]{pasj01}
\usepackage{graphicx,natbib}

\begin{document} 
\Received{}%{yyyy/mm/dd}
\Accepted{}%{yyyy/mm/dd}
%\Published{yyyy/mm/dd}

\title{Galaxy Interactions Trigger Rapid Black Hole Growth: an unprecedented
  view from the Hyper Suprime-Cam Survey }

\author{Andy D. \textsc{Goulding}\altaffilmark{1}}

\altaffiltext{1}{Department of Astrophysical Sciences, Princeton University, Peyton Hall, Princeton, NJ 08544, USA.}
\email{goulding@astro.princeton.edu}
\author{Jenny E. \textsc{Greene}\altaffilmark{1}}
\author{Rachel \textsc{Bezanson}\altaffilmark{1}}
\author{Johnny \textsc{Greco}\altaffilmark{1}}
\author{Sean \textsc{Johnson}\altaffilmark{1}}
\author{Alexie \textsc{Leauthaud}\altaffilmark{2}}
\author{Yoshiki \textsc{Matsuoka}\altaffilmark{3}}
\author{Elinor \textsc{Medezinski}\altaffilmark{1}}
\author{Adrian M. \textsc{Price-Whelan}\altaffilmark{1}}

\altaffiltext{2}{Department of Astronomy and Astrophysics, University of California, Santa Cruz, 1156 High
Street, Santa Cruz, CA 95064, USA}
\altaffiltext{3}{Research Center for Space and Cosmic Evolution, Ehime University, Matsuyama, Ehime 790-8577, Japan}

 %Do NOT move this preamble from here!

\maketitle

\begin{abstract}
  Collisions and interactions between gas-rich galaxies are thought to
  be pivotal stages in their formation and evolution, causing the
  rapid production of new stars, and possibly serving as a mechanism
  for fueling supermassive black holes (BH). Harnessing the exquisite
  spatial resolution ($\sim$0.5 arcsec) afforded by the first
  $\sim$170~deg$^2$ of the Hyper Suprime-Cam (HSC) Survey, we present
  our new constraints on the importance of galaxy-galaxy major mergers
  (1:4) in growing BHs throughout the last $\sim$8 Gyrs. Utilizing
  mid-infrared observations in the WISE All-Sky survey, we robustly
  select active galactic nuclei (AGN) and mass-matched control galaxy
  samples, totaling $\sim$140,000 spectroscopically confirmed systems
  at $i<22$~mag. We identify galaxy interaction signatures using a
  novel machine-learning random forest decision tree technique
  allowing us to select statistically significant samples of
  major-mergers, minor-mergers/irregular-systems, and non-interacting
  galaxies. We use these samples to show that galaxies undergoing
  mergers are a factor $\sim 2-7$ more likely to contain luminous
  obscured AGN than non-interacting galaxies, and this is independent
  of both stellar mass and redshift to $z<0.9$. Furthermore, based on
  our comparison of AGN fractions in mass-matched samples, we
  determine that the most luminous AGN population
  ($L_{\rm AGN} \gtrsim 10^{45}$~erg~s$^{-1}$) systematically reside
  in merging systems over non-interacting galaxies. Our findings show
  that galaxy--galaxy interactions do, on average, trigger luminous
  AGN activity substantially more often than in secularly evolving
  non-interacting galaxies, and we further suggest that the BH growth
  rate may be closely tied to the dynamical time of the merger system.
\end{abstract}

\section{Introduction}

The connections between galaxy--galaxy interactions and the triggering
and/or presence of accreting supermassive black holes (hereafter,
active galactic nuclei; AGN) are a matter of significant on-going
debate. In the broad scope of galaxy evolution, there are many
compelling theoretical reasons to expect a connection between the
encounters of two (or more) gas-rich galaxies with similar (1:$<4-5$)
stellar masses, and the accretion of material onto at least one of the
BHs present in these systems (e.g.,
\citealt{Volonteri:2003aa,Hopkins:2005ad,Di-Matteo:2005aa,Springel:2005aa}). Concurrent
BH growth and the rapid production of new stars (e.g.,
\citealt{Somerville:2008aa,Angles-Alcazar:2013aa}) can naturally give
rise to known correlations between the BH mass and galaxy properties,
such as the bulge mass and stellar velocity dispersion (e.g.,
\citealt{Magorrian:1998aa,Ferrarese:2000aa,Tremaine:2002aa,Gultekin:2009aa,McConnell:2013aa}). Furthermore,
self-regulation of the AGN activity, due to so-called `quasar mode'
feedback processes, can serve as a violent mechanism capable of
disrupting the on-going star-formation by depositing energy back into
the galaxy merger, heating the gas and/or even expelling material out
into the wider dark matter halo. Indeed, AGN feedback is a widely
accepted solution for the formation of massive quiescent early-type
galaxies, and the build-up of the red-sequence, in cosmological
simulations.

A particularly persuasive argument for a connection between the most
rapidly growing BHs and major-mergers is that galaxy interactions
provide a simple solution to the `angular momentum problem'. In
principle, growing a BH requires only a source of cool gas to fuel the
nucleus, supplies of which are typically plentiful in the host
galaxy. However, continuously transporting significant quantities of
this material from the gas reservoirs in the host, down to scales in
which it can accrete onto the BH, while simultaneously dissipating the
specific angular momentum of the gas, is a non trivial issue. Models
of galaxy--galaxy mergers show that tidal forces between the galaxies
can cause gas to be subject to substantial gravitational torques,
resulting in the efficient loss of angular momentum, ultimately
causing substantial gas flow towards the BH, and igniting a powerful
AGN (e.g.,
\citealt{Barnes:1991aa,Mihos:1996aa,Di-Matteo:2008aa,Angles-Alcazar:2017aa}). Despite
the theoretical successes of AGN--galaxy co-evolution models,
observational evidence for a connection between merging galaxies,
galaxy instabilities, and the enhanced presence of AGN activity,
is still inconclusive.

Recent observations of luminous ($L_{\rm AGN} > 10^{46}$~erg~s$^{-1}$)
dust reddened $z \sim 1$--2 quasars have revealed these AGN to be
overwhelmingly hosted by galaxy mergers (e.g.,
\citealt{Urrutia:2008aa,Glikman:2012aa,Glikman:2015aa}), possibly
suggesting that the most luminous BH growth is increasingly likely to
be triggered by galaxy interactions (e.g.,
\citealt{Treister:2012aa,Fan:2016aa}). However, others do not observe
a rise in the incidence of mergers at the highest AGN luminosities
(e.g., \citealt{Schawinski:2012ab,Villforth:2014aa}). Furthermore, at
similar redshifts, AGN with more moderate luminosities
($L_{\rm AGN} \sim 10^{43}-10^{44}$~erg~s$^{-1}$) also appear no more
likely to show interaction signatures than non-AGN systems (e.g.,
\citealt{Cisternas:2011aa,Schawinski:2011aa,Kocevski:2012aa}). Similarly,
at $z<1$, large-scale galaxy interaction signatures such as mergers,
and galaxy-scale bars and instabilities, do not appear to
significantly boost the likelihood of hosting a lower luminosity AGN
($L_{\rm AGN} \lesssim 10^{43}$~erg~s$^{-1}$;
\citealt{Athanassoula:1992ab,Ho:1997aa,Regan:1999ab,Cisternas:2015aa,Cheung:2015aa,Goulding:2017aa}). By
contrast, others find evidence supporting a correlation between
merging and AGN in some more nearby galaxies (e.g.,
\citealt{Koss:2010aa,Ellison:2011ac,Silverman:2011aa,Ellison:2013aa,Satyapal:2014aa,Hong:2015aa,Weston:2017aa}).

Typically, studies have focused on selecting large samples of AGN, and
then comparing the host galaxies of these AGN to non-AGN
systems. However, AGN activity is a stochastic process that is
believed to vary on timescales far shorter than changes related to
galaxy-wide processes (e.g., morphology, star-formation). AGN
variability may therefore cause dilution of would-be significant
correlations between average BH accretion and on-going star-formation
(e.g., \citealt{Chen:2013aa,Hickox:2014aa}) and/or stellar mass
\citep{Yang:2017ab}. Further complicating the observational
view-point, the importance of galaxy interactions for triggering AGN
may also be obscuration dependent, as well as merger-stage dependent
(e.g.,
\citealt{Kocevski:2015aa,Koss:2016aa,Weston:2017aa,Ricci:2017aa}). More
often than not, these previous investigations were hampered by the
ability to sample significant numbers of AGN and mergers that cover
large dynamic ranges in AGN luminosity, and for the more high-redshift
studies, the ability to accurately identify AGN or robustly detect the
presence of galaxy interaction signatures. Progress in the field can
therefore be made by bridging the gap between the low and
high-redshift studies, through the construction of large samples of
merging and non-merging galaxies with deep high-resolution imaging to
$z \sim 1$, which simultaneously encompass statistically significant
populations of moderate to extremely luminous AGN. 

Using dedicated telescopes, wide-format surveys such as the Sloan
Digital Sky Survey (SDSS) carried out comprehensive multi-band imaging
surveys of significant fractions of the sky. These surveys have been
incredibly successful in characterizing the properties of extremely
large galaxy/AGN samples
\citep[e.g.,][]{Strateva:2001aa,Vanden-Berk:2001ab,Strauss:2002aa,Eisenstein:2005aa,Ross:2013aa}. Owing
to mirror size and total integration times these surveys were
necessarily limited to the relatively nearby Universe ($z<0.2$), while
still encompassing large survey volumes of
$V \sim 0.2 (h^{-1}$Gpc)$^3$.  Following in the footsteps of SDSS, the
next generation of wide-format imaging surveys, capable of providing
SDSS-like volumes and imaging quality out to $z=1$ are beginning to
take shape. The on-going Hyper Suprime-Cam (HSC) survey
\citep{Aihara:2017aa} is now providing an unprecedented new view of
the Universe. The combination of the wide field of view and large 8.2
meter mirror diameter provided by the Subaru Telescope gives the HSC
survey exquisite sensitivity and resolving power. Upon completion, the
Wide survey layer of HSC will image $\sim 1400$~deg$^2$ in $grizy$ to
a depth of $i \sim 26$~mag and with a typical $i$-band seeing of
$\sim 0.5''$, less than half that of the median seeing in SDSS
($\sim 1.4''$). Given that the angular diameter increases by only a
factor $\sim 2.5$ from $z=0.2$ to $z=1$, HSC is now allowing the
exploration of galaxy morphologies with SDSS-like precision in
SDSS-like survey volumes out to $z \sim 1$.

Here we harness the unprecedented sensitivity of the first 170 deg$^2$
of the HSC survey combined with complementary all-sky data available
from Wide-field Infrared Survey Explorer (WISE;
\citealt{Wright:2010aa}) to explore the incidence of mid-infrared
(mid-IR; $\lambda \sim 3$--100$\mu$m) identified AGN in merging
galaxies out to $z \sim 0.9$ as a function of the AGN host galaxy
properties. In Section 2 we define our spectroscopic sample of massive
galaxies that have been observed as part of the HSC survey. In Section
3, we outline the modeling of the spectral energy distributions for
the sample to determine their rest-frame photometry and intrinsic
properties, in order to match our galaxy sample in color and stellar
mass, and we utilize the ALLWISE catalog to identify those galaxies
containing luminous AGN. In Section 4, we describe our novel
implementation of a machine learning algorithm to identify interacting
and non-interacting galaxies by harnessing the HSC imaging. In Section
5 we present the incidence of AGN in interacting and non-interacting
galaxies, finding that AGN are, on average, at least a factor
$\gtrsim 3$ more abundant in merging systems, and the most luminous
AGN at fixed stellar mass are preferentially found in merging
galaxies. In Section 6 we discuss the implication of our results, and
outline a framework, linking the observed AGN fractions to the
dynamical time of the merger system. Our concluding remarks are
presented in Section 7. All magnitudes are in the AB system, unless
otherwise stated. Throughout we assume a standard flat $\Lambda$CDM
cosmology with $H_0 = 70$~km~s$^{-1}$~Mpc$^{-1}$ and
$\Omega_{\rm M} = 0.3$.

\begin{figure}
%  \begin{center}
  \hspace{0.5cm}
    \includegraphics[width=0.95\linewidth]{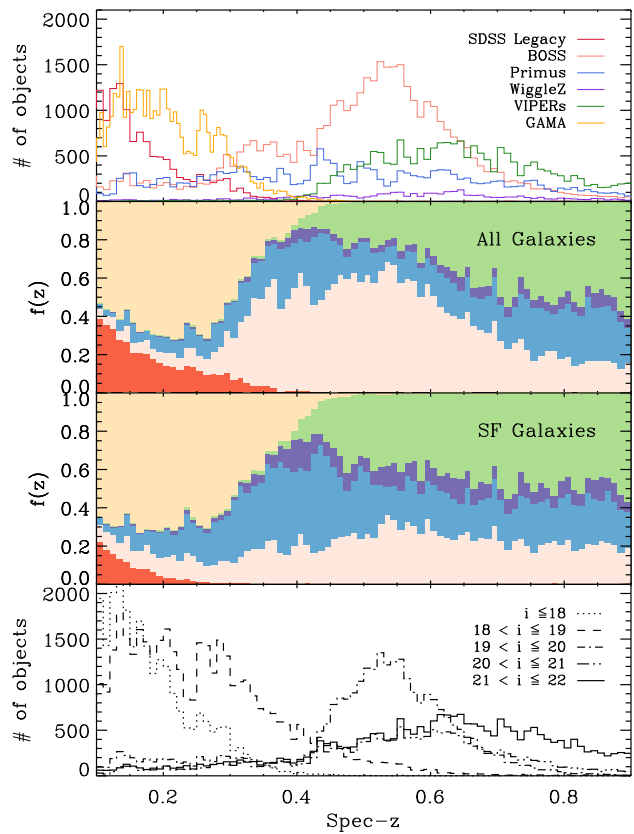}
    \vspace{0.5cm}
%  \end{center}
    \caption{{\bf Top:} Distribution of spectroscopically identified
      galaxies in our parent sample as a function of spectroscopic
      survey (SDSS-II Legacy, SDSS-III BOSS, PRIMUS, VIPERs, GAMA and
      Wiggle-Z). {\bf Middle:} Fractional contribution of a given
      spectroscopic survey to our parent galaxy sample as a function
      of redshift for all galaxies within our parent sample, and for
      the star-forming galaxy sample after applying the UVJ cut shown
      in Fig~\ref{fig:uvj}.  {\bf Bottom:} Spectroscopic sample as a
      function of $i$-band magnitude.}\label{fig:fig1}
\end{figure}

\section{Sample Selection in the HSC Survey}

\subsection{The Hyper Suprime-Cam Survey}

The Hyper Suprime-Cam (HSC) Survey is an ambitious 300 night imaging
survey undertaken as part of the Subaru Strategic Program (SSP;
\citealt{Aihara:2017aa}). The HSC survey is designed to provide nested
wide-field multi-band imaging over a total of $\sim 1400$~deg$^2$
using the HSC instrument on the Subaru 8.2m telescope on Mauna Kea in
Hawai’i. HSC is constructed of 116 (104 science detectors) Hamamatsu
Deep Depletion 2K$\times$4K CCDs, with a 1.77 deg$^2$ field-of-view
(FOV) and has an instrumental Point-Spread Function (PSF) of
$D_{\rm 80}<$0.2$''$ (80\% enclosed light fraction) over the entire
FOV across all imaging filters.

The HSC survey consists of three survey layers: the {\it Wide} layer
covers a solid angle of 1400~deg$^{2}$ in $grizy$ filters to a depth
of $r \approx 26$~mags (5$\sigma$, point source). The 27 deg$^2$ {\it
  Deep} layer reaches $r \approx 27$~mags, with the addition of three
narrow-band filters at $\lambda \sim 3870$,8160 and 9210\AA, and the
3.5~deg$^2$ {\it Ultradeep} layer is a further $\approx$1~mag fainter
than {\it Deep}, allowing detection of Ly$\alpha$ emitters to
$z \approx 7$. In addition, the HSC Survey fields were carefully
constructed to overlap with existing multi-wavelength survey fields,
e.g., millimeter data from the Atacama Cosmology Telescope (ACT);
X-ray data from multiple XMM-Newton and Chandra; near-/mid-infrared
imaging surveys such as VIKING/VIDEO; UKIDSS; {\it Spitzer} and {\it
  WISE}; and optical spectroscopic surveys such as SDSS Legacy/BOSS,
PRIMUS \citep{Coil:2011ab}, VIPERS \citep{Guzzo:2014aa}, GAMA
(\citep{Liske:2015aa}, Wiggle-Z \citep{Drinkwater:2010aa}, COSMOS
\citep{lilly09}, and DEEP2 \citep{Newman:2013aa}. For the
  specific region centers of the three layers that make up the HSC-SSP
  survey, we refer the reader to \citealt{Aihara:2017aa}.

The data used throughout this manuscript is based on an internal
release of the {\it Wide} layer data, release S16A, and covers
$\sim 170$~deg$^2$. Basic data processing, including bias and
background subtraction, flat-fielding, astrometric calibration,
individual exposure co-addition, and object detection was performed
using {\tt hscPipe v4.0.1}, which is an HSC-specific derivative of the
Large Synoptic Survey Telescope (LSST) processing pipeline. For
further details regarding {\tt hscPipe} and the HSC-SSP data releases,
see Bosch et al. (2017) and \cite{Aihara:2017ab}. For our analyses we
use the source catalogs, images, and other relevant data derived from
the co-added HSC images produced by {\tt hscPipe}. The image co-adds
are shifted to a common World Co-ordinate System (WCS), and have a
pixel scale of $0.168''$.

\subsection{Outline of this manuscript}

Our primary goal is to constrain the effects of gas-rich merging of
galaxies on the growth of BHs out to $z \lesssim 1$. To achieve this
we require:

\begin{enumerate}

\item deep, high spatial resolution optical imaging (from HSC) of a
  large parent sample of galaxies spectroscopically confirmed to be in
  the redshift range $0.1 < z < 0.9$, from which we can robustly
  identify large sub samples of interacting and non-interacting
  galaxies. See Section~\ref{sec:speczsample}.

\item the intrinsic properties of the parent galaxy sample, such as
  rest-frame colors, stellar masses, star-formation rates, derived
  using SED fitting (see Section~\ref{sec:sedfitting}). These
  measurements allow us to perform property-matched tests between
  different interaction-state systems, such as stellar mass matching,
  and rest-frame color matching, using diagnostics such as the UVJ
  diagram (see Section~\ref{sec:uvj}).

\item a homogeneous obscuration independent selection of AGN within
  our parent sample. We achieve this through the use of mid-infrared
  color diagnostics performed on photometry obtained from the WISE
  All-sky survey. See Section~\ref{sec:agnID}.

\item an accurate automated method for classifying signatures of
  recent/on-going merger events. We achieve this through a novel
  implementation of a Random Forest machine learning algorithm,
  trained on a large sample of visually identified mergers and
  non-mergers within the HSC $i$-band imaging. See
  section~\ref{sec:randomforest}.

\end{enumerate}

In Section~\ref{sec:results}, we present the results of our
investigation, and show conclusive statistical evidence that BHs
hosted by merging galaxies are at least three times more likely to be
rapidly growing at high-Eddington ratios than a mass-matched sample of
non-interacting galaxies. This suggests not only that AGN are
triggered by merging, but also the rapid growth of the BH(s) can be
sustained during the merger event.

\subsection{Selecting bright ($i<22$~mag) galaxies in HSC} 
\label{sec:speczsample}

In this section we describe our selection techniques in order to
construct samples of interacting and non-interacting galaxies with
firm spectroscopic redshifts. In Section~\ref{sec:results}, we use
these galaxy samples to assess the importance of galaxy interactions
on the growth of BHs to $z<1$. Our parent galaxy sample contains all
objects with $i<22.3$~Kron magnitudes in the S16A data release of the
HSC survey. We set the {\tt detect\_is\_tract\_inner}, {\tt
  detect\_is\_patch\_inner}, {\tt detect\_is\_primary} and {\tt
  is\_extended} data flags on the sample, as we require the most
complete galaxy sample available within the HSC database\footnote{We
  do not make use of the HSC Bright Star Masks. Due to the
  substantially poorer PSF in WISE, galaxies with HSC photometry that
  are contaminated by foreground stars are typically not identified
  in WISE, and hence our excluded during our HSC--WISE cross-match
  outlined in Section~\ref{sec:agnID}. Furthermore, the inclusion of
  the HSC Bright Star Mask would potentially mask bright/large
  galaxies, which would bias our sample against the most massive
  merging systems.}. Our choice of flux limit derives from our ability
to (1) recover the source morphologies and robustly identify
interacting galaxy features such as disturbances, irregular
morphologies, tidal tails and bridges out to $z < 0.9$ (a detailed
analysis of the HSC galaxy morphologies and comparison to Hubble Space
Telescope data will be presented in a future publication; Goulding et
al. in prep.), (2) the completeness of spectroscopic catalogs within
the survey regions; and (3) the addition of a systematic uncertainty
of $\pm 0.3$~mags due to difficulty in measuring photometry of merging
systems.

Each source extracted from the HSC database is then cross-matched to
within $<1''$ with the publicly available spectroscopic redshift
(spec-$z$) catalogs pertaining to the survey sky-regions. The median
separation between the HSC and the spec-$z$ position is $\sim 0.13''$.
Specifically, we harness spec-$z$ measurements from the SDSS Legacy
Catalog (complete to $r < 17.77$), the SDSS-DR12 BOSS survey
(color-selected galaxies, and approximately stellar mass limited; see
\citealt{Dawson:2013aa,Maraston:2013aa,Reid:2016aa,Leauthaud:2016aa}),
the GAMA-DR2 survey (complete to $r < 19.0$), the PRIMUS survey
(complete to $i<22.5$; Coil et al 2011), the WiggleZ Dark Energy
survey ($20.0<r<22.5$; Drinkwater et al. 2009) and the first data
release of the VIMOS Public Extragalactic Redshift Survey (VIPERs;
$i < 22.5$; \citealt{Garilli:2014aa,Guzzo:2014aa}).

Our requirement of a detected $i<22.3$~mag source in HSC-S16A WIDE, as
well as a publicly available spectroscopic redshift (within at least
one of the aforementioned surveys), and HSC imaging with a well
characterized point spread function, results in a combined area of
$\sim 170$~deg$^2$, and a galaxy catalog containing 140,158 unique
galaxies at $0.1 < z < 0.9$.  Similar to our brightness limit, the
imposed redshift limits are based on the bright photometry limit for
HSC \citep[see][]{Bosch:2017aa}, the targeted completeness limits for the
spectroscopic redshift surveys, and our ability to accurately
determine the morphological classifications, and identify low-surface
brightness tidal-tails of the systems from the WIDE-depth (i$<$25.9
mag) HSC imaging at higher redshifts.

In Figure~\ref{fig:fig1} we present the breakdown of the spectroscopic
redshift distributions for our parent sample as a function of the
redshift survey from which the spec-$z$ originated, and as function of
the source brightness in the $i$-band. It is clear that at any given
redshift our total sample is dominated by 2--3 of the redshift
surveys. For example, at lower redshifts ($z< 0.3$), our sample is
mainly composed of objects selected from the SDSS-Legacy and GAMA
surveys, while at redshifts $0.4 < z< 0.6$, the sample is driven by
SDSS-BOSS systems, with roughly equal sub-dominant contributions from
VIPERs and PRIMUS galaxies.\footnote{We note that the optical
  color-selection imposed as part of the BOSS survey, in order to
  select a galaxy for spectroscopic targeting, was designed to
  identify massive, passively evolving systems. In
  Section~\ref{sec:agnID}, we find that $< 4$\% of AGN in our sample
  are hosted by quiescent systems. As these passively evolving
  galaxies do not appear to be contributing significantly to the
  overall growth of BHs, we choose to remove these particular objects
  from our spec-$z$ sample. Removing the quiescent BOSS objects
  mitigates this strong selection bias within the redshift range
  $0.4 < z< 0.6$, and we defer the investigation of `dry mergers' to
  future studies. } While our sample is by selection heterogeneous,
and contains a range of selection functions for the different surveys,
we will demonstrate directly in Section~\ref{sec:obsbias} that our
results are not sensitive to the details of the different samples.

\begin{figure*}
  \begin{center}
    \includegraphics[width=\textwidth]{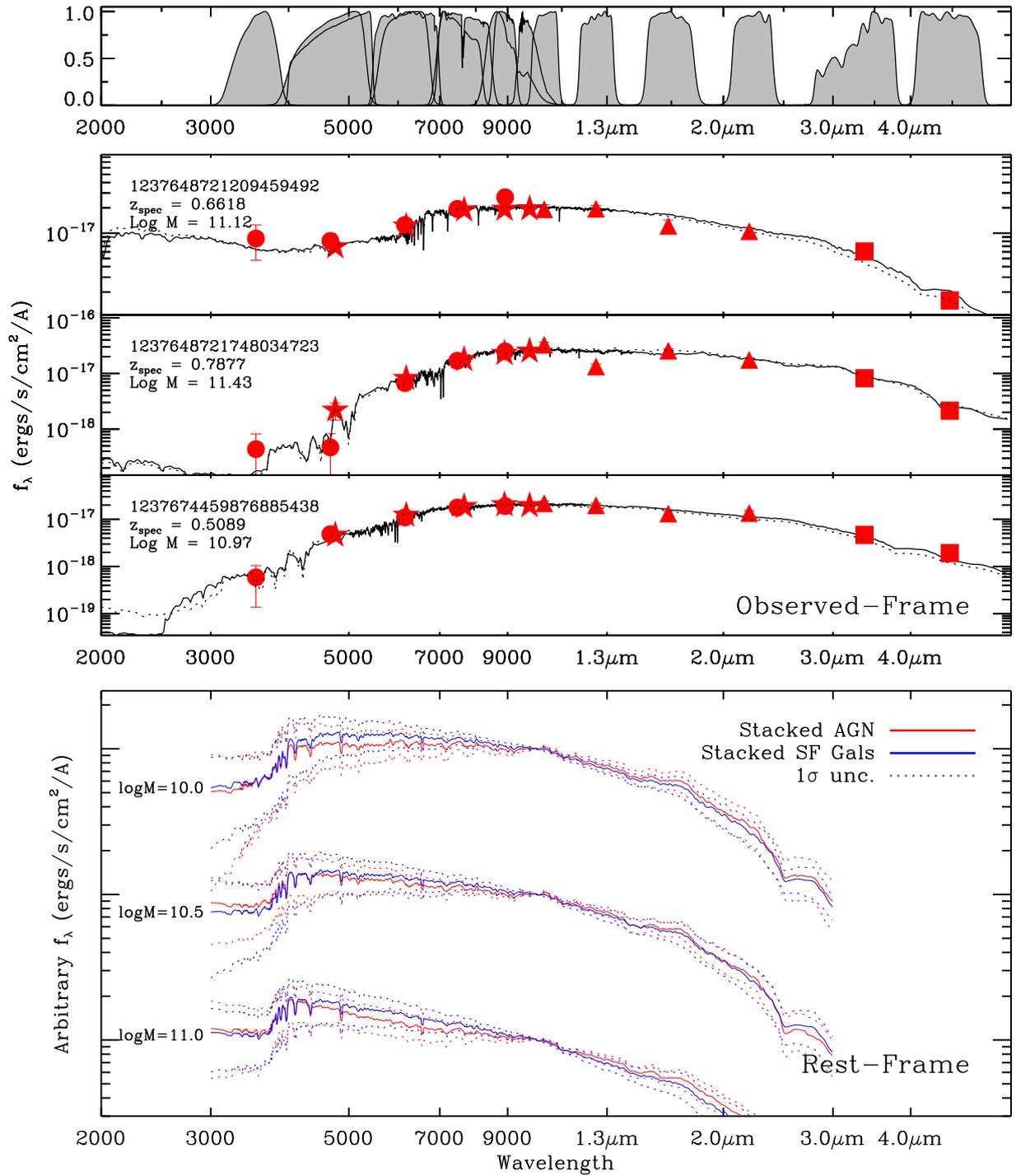} 
  \end{center}
  \caption{Examples of FAST produced spectral energy distribution
    (SED) fits while harnessing available photometry SDSS ($ugriz$;
    filled circles), HSC ($grizy$; filled stars), UKIDSS LAS $YJHK$
    (filled triangles) and WISE ($W1W2$; filled squares) and spectral
    redshifts. SED fits are shown with (solid line) and without
    (dashed line) the inclusion of the HSC $grizy$ photometry. Top
    panel provides the normalized filter response used throughout our
    analysis. Lower panel provides the stacked rest-frame SEDs (solid
    lines) and the 1-$\sigma$ spread in the SED models (dotted-lines)
    for the obscured-AGN (red) and the star-forming galaxies (blue) in
    three bins of stellar mass: log$(M_*/M_{\odot}) \sim 10$,10.5 and
    11.0.}\label{fig:sed}
\end{figure*}

\section{Sample Properties}

\subsection{SED modeling using FAST}
\label{sec:sedfitting}

In this section, we use a suite of available photometry, in
conjunction with the spectroscopic redshifts to derive the UV to IR
spectral energy distributions (SEDs) for our galaxies in our spec-$z$
sample, defined in Section~\ref{sec:speczsample}. We use SED modeling
to derive physical properties such as the stellar mass ($M_*$),
star-formation rate (SFR), and dust extinction ($A_V$). We use these
derived measurements in Section~\ref{sec:results} to produce
$M_*$-matched samples of interacting and non-interacting galaxies.

In order to accurately constrain the galaxy light blueward of the
4000$\AA$ break, for objects in our lowest redshift bin $z<0.3$, we
require $u$-band photometric measurements. For this purpose we choose
to harness the available Petrosian magnitude $ugriz$ photometry from
the 12th data release (DR12) of the SDSS survey, which is complete to
$r \lesssim 22.4$~mags. All of the galaxies in our main spec-$z$
galaxy sample are detected in at least the $g$, $r$ and $i$ bands in
SDSS-DR12. At faint magnitudes ($i_{\rm SDSS,Petro}>20.5$), the
uncertainties in the SDSS photometry become large, due to the
sensitivity limit of the SDSS observations. Given the depth of the HSC
data, the inclusion of HSC photometry will serve to increase the
precision of our SED modeling for faint systems. Hence, for sources
with $i_{\rm SDSS,Petro}>20.5$ magnitudes, we supplement the
observed-frame optical data with HSC $grizy$ Kron-magnitude
photometry.

Typically, extended sources with $i \gtrsim 20.5$ mags have HSC
$i$-band photometry that is consistent ($\pm 0.05$~mags) with the
photometry from SDSS. However, some sources with
$i_{\rm SDSS,Petro}>20.5$~mag still have significantly discrepant
photometric measurements between SDSS and HSC
($|i_{\rm HSC,kron} - i_{\rm SDSS,Petro}| > 0.15$~mags). Such a
difference between the SDSS and HSC photometry is well beyond the
typical statistical uncertainty quoted for the photometry in either
survey ($\sigma_{\rm SDSS} \sim 0.09$; $\sigma_{\rm HSC} \sim 0.04$),
and points towards a photometric measurement issue for a given object
in the HSC pipeline.\footnote{At the present time, galaxies observed
  to have $i \sim 20.5$ in SDSS are subject to significant shredding
  ($\sim 20$\% of disk galaxies) and deblending issues in HSC, as well
  as having over-subtracted backgrounds in the HSC imaging
  \citep{Aihara:2017ab,Bosch:2017aa}). Presently, for bright disk
  galaxies with $i < 19$~mags, only $\sim 20$\% show good photometric
  agreement between SDSS and HSC in the $i$-band (at the
  $\pm 0.05$~mag level), with $\sim 55\%$ having over-estimated
  photometry in HSC in the range 0.1--0.7 mags. Furthermore, at
  $i<18$~mags, HSC imaging is often subject to saturation towards
  galaxy centers and at the positions of bright star clusters, which
  necessitates the use of the SDSS photometric measurements.} Hence,
we choose not to include the additional HSC photometry for these
objects. We note that the exclusion of the HSC photometry for some
objects does not significantly affect our SED fitting procedure or its
derived measurements, as we show in Figure~\ref{fig:sed}.

To place more stringent constraints on the stellar mass of our galaxy
sample, it is prudent to measure the stellar-light centered around the
rest-frame near infrared, hence we include available $YJHK$ photometry
from the Deep Extragalactic Survey (DES) and the Large Area Survey
(LAS), which are part of the 10th data release of the UKIRT InfraRed
Deep Sky Surveys (UKIDSS), which covers the spec-$z$ surveys
considered throughout this manuscript. Specifically, we use the
Petrosian magnitudes available within UKIDSS-DES and LAS, which we
correct for aperture biases between SDSS/HSC and UKIDSS. Based upon
our comparison of the UKIDSS-DR10 catalog photometry with our own
SDSS/HSC aperture-matched photometry, which we extracted directly from
the UKIDSS imaging for a random subset of the sources in our spec-$z$
catalog, we determined that a flux-dependent correction to the UKIDSS
Petro photometry of $+0.02-0.05$ magnitudes, produces adequately
matched photometric measurements between the three catalogs. We
further confirmed our aperture corrections by harnessing the
aperture-matched catalog of GAMA/SDSS and UKIDSS sources
\citep{Hill:2011aa}, finding a similar systematic average offset
between the UKIDSS-DR10 Petro measurements and GAMA/SDSS--UKIDSS
Petro-mags of $\sim+0.03$~mags.

At higher redshifts ($z>0.65$), the rest-frame near-IR moves into the
mid-IR, hence, we also include the 4-band $W1$--4 mid-infrared
photometry from the WISE All-Sky Survey where AGN emission may also be
prominent. In Section~\ref{sec:agnID} we use the WISE mid-IR
photometry to build our AGN sub-samples using WISE color--color
diagnostics.

We use the publicly available {\sc idl} code, FAST
\citep{Kriek:2011aa} to model the optical--IR SED of each object to
derive physical properties, as well as the appropriate k-corrections
required to produce rest-frame photometry for each galaxy. FAST
searches over a grid of models and uses $\chi^2$-statistics to
determine the best solution. Throughout the fitting we assume an
exponentially declining star-formation history with
SFR$\sim {\rm exp}[-t/ \tau]$ and a characteristic time-scale of
log$(\tau) = 7.0-10.0$~yr, a Chabrier initial mass function, assuming
stellar ages in the range, log$({\rm age}) = 8-10.1$, and the
high-resolution stellar population synthesis models of
\cite{Bruzual:2003aa}. Furthermore, we use a \cite{Calzetti:2000aa}
dust reddening curve, and allow for extinction in the range
$A_V = 0.2-4.0$ and derive templates for metalicities of
$\{0.008,0.02({\rm solar}),0.05\}$. To determine the uncertainties for
the fitted parameters we perform 500 Monte Carlo realizations of this
FAST setup, and quote the 67th percentile of the simulations.

The Bruzual \& Charlot SPS templates do not include contributions from
AGN. Hence, to further ensure that we do not overestimate the stellar
mass of sources containing a mid-IR detected AGN, for known AGN (see
Section~\ref{sec:agnID}) we fit the WISE photometry with a power-law,
and following \cite{Azadi:2017aa}, we subtract this AGN continuum from
the WISE photometry to estimate the galaxy-only continuum. In
Figure~\ref{fig:sed} we show examples of the two best-fit SED
templates to our suite of photometry assuming the inclusion (solid
line) or exclusion (dotted line) of the HSC photometry.\footnote{AGN
  continuum emission in Type-1 systems will also contribute to the
  photometry at rest-frame $\lambda \lesssim 4000$\AA resulting in
  poorly determined galaxy properties during the SED fitting
  process. However, we robustly identify and remove Type-1 AGN from
  our sample in Section~\ref{sec:agnID}.} Qualitatively, the best-fit
templates appear extremely similar across a large wavelength range
($\lambda \sim 3000$\AA--4$\mu m$). Indeed, we find that the
difference between the derived $M_*$, SFR and $A_V$ measurements
between the two photometry sets are all consistent at the 1$\sigma$
uncertainty level determined directly from our Monte Carlo
realizations. In turn, this suggests that for the subset of
significantly extended and/or well-resolved galaxies currently lacking
reliable photometry in HSC, the exclusion of these photometric points
does not affect our ability to measure the galaxy properties using
FAST. Furthermore, we assessed systematic bias towards the $M_*$
measurements between the (obscured) AGN and non-AGN galaxies. In the
lower panel of Fig.~\ref{fig:sed} we provide the rest-frame stacked
SED templates for sources predicted to have
$M_* \sim 10^{10},~10^{10.5}$, and $10^{11}~M_{\odot}$ for AGN and
non-AGN. We show that in each instance that the stacked SED templates
are similar for AGN and non-AGN systems in each individual mass
bin. This suggests that scattered light from the obscured AGN is not
present or not adversely affecting the stellar mass estimates in these
systems. Hence, we conclude that there is no significant systematic
bias between the stellar mass estimates for AGN and non-AGN galaxies.

\subsection{Rest-frame photometry and UVJ selection}
\label{sec:uvj}

\begin{figure}
  \begin{center}
    \includegraphics[width=\linewidth]{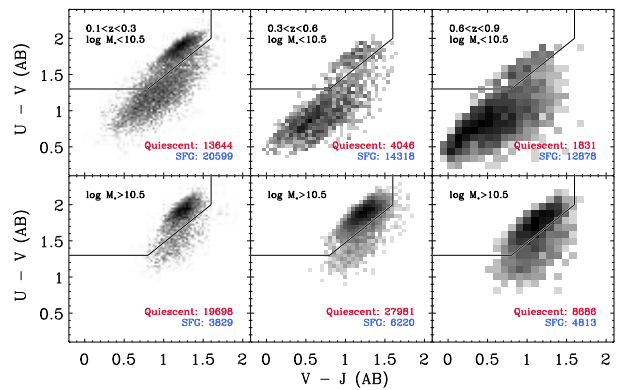} 
  \end{center}
  \caption{Rest-frame U--V versus V--J diagrams for all spec-z galaxies
    in our parent sample. Rest-frame AB photometry is derived from
    best-fit SED templates produced from FAST. Column panels: galaxy
    sample split in three redshift bins, $0.1<z<0.3$, $0.3<z<0.6$ and
    $0.6<z<0.9$. Rows: redshift bins separated by two stellar mass
    bins, log$M_*/M_{\odot} < 10.5$ and
    log$M_*/M_{\odot} \gtsim 10.5$.}\label{fig:uvj}
\end{figure}

As well as matching our interacting and non-interacting galaxy samples
based on their stellar masses, it is also prudent to consider matching
on galaxy color. In this section we use the derived rest-frame
photometric information to separate our spec-$z$ galaxy sample using a
typical star-forming/quiescent classification diagram, which harnesses
the apparent rest-frame color bi-modality between star-forming and
non-star-forming galaxies.

Following the procedure outlined in the previous section, we determine
rest-frame photometry directly from the best-fit SPS template. We
apply a simple Gaussian noise model to the best-fit template with the
noise amplitude matched to the average 1-$\sigma$ uncertainty of the
measured photometric data, and use the known spec-$z$ to produce a
rest-frame simulated SED. The addition of the Gaussian noise provides
a more realistic estimate of the measurement of the photometry that
would be found from real observations, and that may otherwise not be
captured in the discretized SPS models. We convolve the simulated SED
with rest-frame $U$, $V$ and $J$ filters, as well as determining
appropriate K-corrections for all of the input photometry, to produce
rest-frame measurements.

In Figure~\ref{fig:uvj}, we show the rest-frame $U-V$ vs $V-J$
color-color diagram (hereafter, UVJ diagram) for our spec-$z$ galaxy
sample. We separate quiescent and star-forming (SF) galaxies using the
proposed boundaries computed by \cite{Williams:2009aa}. The
implementation of this two color cut allows us to identify
dust-reddened star-forming galaxies, i.e., those with $U - V > 1.6$
that may otherwise be tagged as quiescent using typical rest-frame
color-magnitude diagrams. We confirm that in three separate and
distinct redshift bins, $z\sim0.1$--0.3, 0.3--0.6 and 0.6--0.9, that
the SF galaxies are typically less massive systems (with
$M_* < 3 \times 10^{10}~M_{\odot}$) than their quiescent
counterparts. At these lower masses, our sample is over-whelmingly
dominated by SF systems by factors of $\sim$6--10 at $z>0.3$, while in
the lowest redshift bin, the samples of quiescent and SF galaxies are
more comparable, owing mainly to the spectroscopic coverage from
GAMA-DR2. Of the more massive galaxies with
$M_* > 3 \times 10^{10}~M_{\odot}$, our spec-z sample is dominated by
quiescent galaxies by factors of $\sim$2--5, driven mainly by the
abundance of massive `red' galaxies targeted for spectroscopic
follow-up in SDSS-BOSS. The more massive population of SF galaxies,
that reside outside of the quiescent boundary in the UVJ diagram, have
systematically higher $U-V$ and $V-J$ colors, than the lower-mass
star-forming galaxies. This suggests at least 1 magnitude of optical
extinction towards the galaxy continuum in these objects. Our final
spec-$z$ sample contains 75,886 and 62,657 quiescent and SF galaxies,
respectively.

\subsection{Mid-IR AGN Selection using the (ALLWISE) WISE All-Sky Survey}
\label{sec:agnID}

While most AGN are intrinsically luminous in any given wavelength
band, the homogeneous selection of an unbiased population of AGN from
survey data is not straightforward. Intervening gas and dust, as well
as dilution of the AGN signatures by host galaxy light, are the most
common sources of observation selection bias. Indeed, many studies
have now shown that no one wave-band can identify all AGN
\citep{Alexander:2008aa,donley08,Hickox:2009aa,Juneau:2011aa,Mendez:2013aa,Goulding:2014aa,Trump:2015aa,Azadi:2017aa}. Moreover,
in the previous section we determined that the most massive SF
galaxies in our spec-$z$ sample are likely obscured by $A_V > 1$~mags,
further hampering AGN detections. However, even in the presence of
significant dust attenuation, relatively unbiased detections of
luminous AGN may be made at mid-IR wavelengths. AGN emission produced
directly from the optical/UV luminous accretion disk or from the X-ray
emitting corona may be absorbed and reprocessed by dust which
surrounds the central BH. This dust-rich torus isotropically re-emits
at mid-IR wavelengths, which is relatively insensitive to further
absorption at larger radial distances from the AGN.

\begin{figure*}
  \begin{center}
    \includegraphics[width=\textwidth]{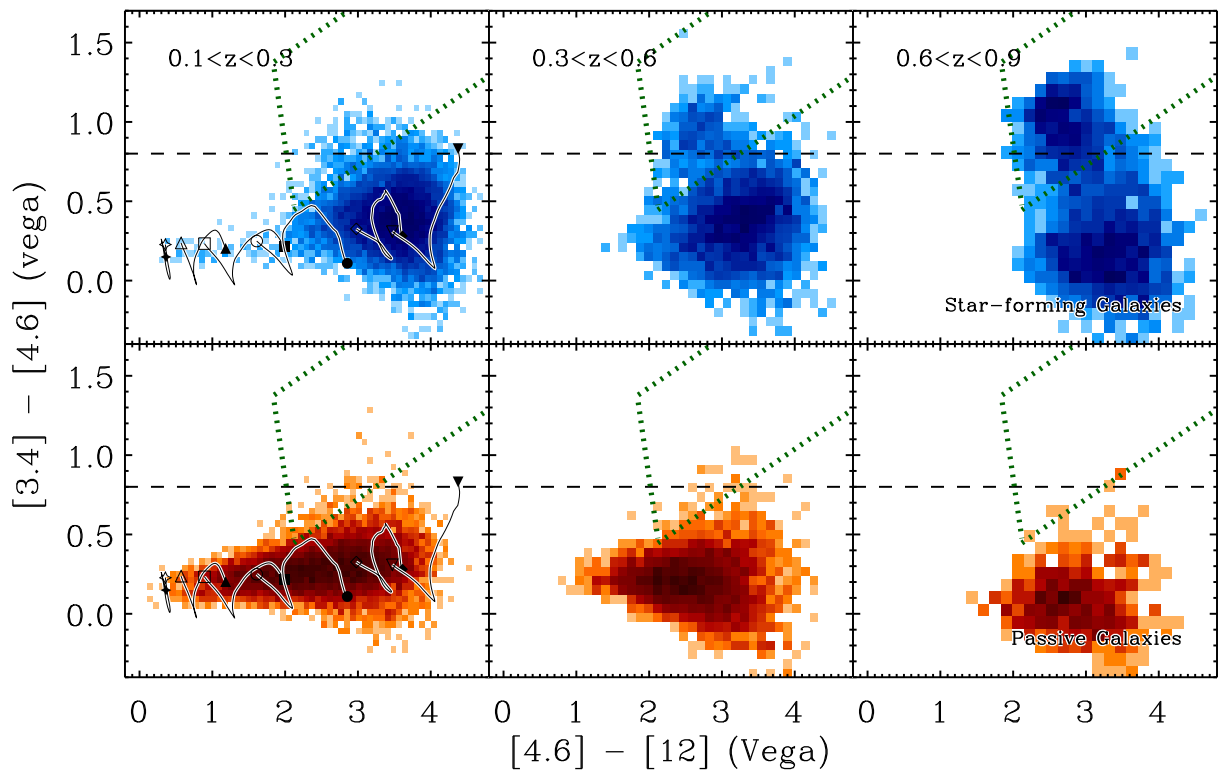} 
  \end{center}
  \caption{WISE infrared [3.4]--[4.6] versus [4.6]--[12.0] Vega
    magnitude color-color diagram of HSC galaxies with WISE
    counterparts. All objects have detections in the [3.4] and [4.6]
    bands with S/N$\gtrsim 4$ and in the [12.0] band with
    S/N$\gtrsim 2.5$. Green dotted box shows the 2-color AGN selection
    region of Mateos et al. (2012) and black dashed line the 1-color
    AGN selection cut of Stern et al. (2012). Panels columns are split
    by redshift, using the same cuts as in Fig.~\ref{fig:uvj}, rows
    indicate galaxies separated according to their position in the UVJ
    diagram. In the $0.1<z<0.3$ panels, simulated color-color tracks
    derived from SED templates (Polletta et al. 2007) are shown. These
    tracks begin at $z=0.1$ (filled symbols) and finish at $z=0.9$
    (open symbols). Individual templates are for a 13~Gyr Elliptical
    (star), a ULIRG (downward triangle), a starburst (diamond), and S0
    (upward triangle), Sb (square) and Sd (circle) spiral
    galaxies.}\label{fig:wise_color_color}
\end{figure*}

Mid-IR AGN identifications can be made either through the detection of
high-ionization emission lines in mid-IR spectroscopy (e.g.,
\citealt{diamond09,ga09}), or through the establishment of the
presence of an AGN-produced powerlaw continuum using mid-IR
photometry. More specifically, a wide-variety of AGN selection
techniques have been proposed that harness color-color diagrams
developed using 2, 3 or 4-band mid-IR photometry taken using the IRAC
instrument on the NASA Spitzer Space Telescope (e.g.,
\citealt{Lacy:2004aa,Stern:2005aa,Alonso-Herrero:2006aa,Donley:2012aa},
or more recently using WISE (e.g.,
\citealt{Jarrett:2011aa,Stern:2012aa,Mateos:2012aa}), which we harness
here. Mid-IR color-color selection is particularly effective at
identifying high-luminosity AGN
($L_{\rm AGN} \gtrsim 10^{43}$~erg~s$^{-1}$), where the contrast
between the AGN and the host-galaxy is high. However, IR color-color
diagrams may fail to readily identify AGN accreting at low Eddington
ratios (e.g., \citealt{Donley:2012aa,Mateos:2012aa,Hainline:2016aa}).

\subsubsection{Identifying WISE counterparts to galaxies in HSC}

Due to the all-sky nature of the WISE survey, HSC is covered in its
entirety by 4-band cryogenic mid-IR observations at 3.4, 4.6, 12 and
22$\mu m$. We matched the positions of our HSC spec-$z$ sample to the
objects present in the the ALLWISE release using the same method
outlined in \cite{dabrusco13} and \cite{Goulding:2014aa}, which we
briefly outline here. We used expanding radial apertures of
$\Delta r = 0.1''$ to search for all WISE counterparts to HSC sources
to a distance of $r = 4''$. By randomly shifting the centroid of the
HSC source by $\pm 20 ''$, we computed the likelihood of spurious
counterparts as a function of radial distance from the true HSC source
position. We determined that the optimal maximum matching radius for
HSC and WISE is $r(<R) \sim 1.6''$. At $r>1.6''$ the probability of
including spurious counterparts into our sample exceeds that of a real
HSC--WISE match.

We identified 103,406 HSC galaxies in our spec-$z$ sample that have at
least one counterpart in WISE. For the $\sim 0.6$\% of HSC sources
with multiple WISE counterparts within 1.6$''$, we chose the WISE
source with the smallest separation to the HSC galaxy. Over 97\% of
the HSC--WISE matches are at separations of $r < 1''$, and the
distribution of the matching radii are characterized by a log-normal,
peaked at 0.18$''$, and with a FWHM of $\sim 0.35$~dex. The peak at
0.18$''$ is consistent with the astrometric precision of the WISE
sources (Wright et al. 2010).

\subsubsection{AGN identification using mid-IR color selection}

In Figure~\ref{fig:wise_color_color} we use the [3.4]--[4.6],
[4.6]--[12.0] mid-IR color-color diagram to identify galaxies with a
significant contribution from a central AGN. Specifically, we include
all galaxies that are detected in the WISE [3.4], [4.6] and [12.0]
bands, with a signal-to-noise (S/N) $\gtrsim 4$ for the [3.4] and
[4.6] bands. The longer wavelength bands in WISE are significantly
less sensitive (by at least a factor 2) than the [3.4] and [4.6]
bands. Hence, we marginally relax our S/N threshold to S/N$>2.5$ in
order to consider a source detected in the [12.0] band. We note this
cut is still more conservative than the S/N$> 2$ used to identify
objects throughout the ALLWISE catalog. 41,990 galaxies in our HSC
spec-$z$ sample are detected in 3-bands using WISE with our S/N cuts,
and an additional 44,389 galaxies are detected in only [3.4] and [4.6]
WISE bands.

We use the two-color IR-AGN wedge defined by \cite{Mateos:2012aa} to
identify [3.4], [4.6] and [12.0] detected objects that have
powerlaw-like continua, indicative of the presence of a
radiatively-efficient AGN. For galaxies that are not detected in the 3
WISE bands considered in Fig.~\ref{fig:wise_color_color}, we also use
the single color cut of \cite{Stern:2012aa} to identify additional AGN
(dashed line in Fig~\ref{fig:wise_color_color}). This has the
advantage of allowing us to boost source statistics due to the
relative insensitivity of the longer wavelength WISE bands, as well as
including the abundance of heavily obscured AGN that reside in
Ultra-Luminous IR galaxies with [4.6]--[12.0]$>3.5$ that may otherwise
be excluded by the 2-color wedge. Mid-IR AGN selections are suspected
to be contaminated by low-metallicity strongly star-forming dwarf
galaxies in the low-redshift universe (e.g., Hainline et al. 2016),
and by hot strongly dust-obscured galaxies beyond $z>2$. In
Fig.~\ref{fig:wise_color_color}, we additionally show simulated
color-color tracks that are derived from the SED templates of Polletta
et al. (2007) in our considered redshift range. These tracks typically
lie below or outside of the AGN selection regions used throughout, and
hence, contamination to our AGN selection from non-AGN interlopers is
likely to be minimal (see also Goulding et al. 2014).

Of the $\sim$41,990 galaxies in our matched HSC--WISE 3-band sample,
3,125 are selected as AGN using the Mateos et al. (2012) WISE
selection method. An additional 665 galaxies are selected as AGN using
the Stern et al. single color cut i.e., a total of 3,790 WISE-selected
AGN. We find that if we separate the galaxies based on their position
in the UVJ diagram, the WISE-AGN are over-whelmingly hosted in
star-forming galaxies. Indeed, only $\sim$4\% of the WISE-AGN in our
spec-$z$ sample are hosted in quiescent galaxies. This clear
separation of mid-IR AGN residing in star-forming galaxies over
quiescent galaxies serves to highlight the previously observed
connection between star-formation rate and BH accretion rate (e.g.,
\citealt{Chen:2013aa,Hickox:2014aa}).

Unlike more traditional color--magnitude diagrams
\citep{Strateva:2001aa,Baldry:2004aa}, the separation of star-forming
and quiescent galaxies through UVJ diagnostics is relatively
insensitive to dust extinction in the host. As a result, dusty
star-forming galaxies are still robustly identified using UVJ while
they may otherwise be classified as quiescent or green-valley systems
using color--magnitude diagrams. The observed separation of AGN in
Fig.~\ref{fig:wise_color_color} using UVJ may explain why many AGN
have previously been believed to be an interesting population of
transitioning `green-valley' galaxies that lie between the blue cloud
and red sequence
\citep{Bell:2004aa,Faber:2007aa,Nandra:2007aa,Hasinger:2008aa,Silverman:2008aa,Mendez:2013aa}. In
reality, it would appear that these luminous mid-IR AGN are merely
hosted in dusty star-forming systems with reddened optical colors.

The lack of mid-IR AGN observed in quiescent galaxies does not suggest
that there are no accreting BHs in these systems. The vast majority of
radio-loud AGN are known to be hosted in massive quiescent galaxies
(e.g.,
\citealt{Best:2005aa,Hickox:2009aa,Goulding:2014aa,Delvecchio:2017aa}),
though the majority of these radio AGN lack the signatures of a
radiatively efficient accretion disk, which would be observed in the
mid-IR. Furthermore, in galaxy group or cluster environments, evidence
of AGN feedback due to radio emission from the BH present in the
brightest cluster galaxy (so called, maintenance mode feedback) has
long been established (e.g.,
\citealt{Best:2005aa,Rafferty:2006aa,McNamara:2007aa,Kauffmann:2008aa,fabian12}). In
these systems, powerful radio lobes inject mechanical energy back into
the intracluster medium, which in turn prevents the efficient cooling
gas, and are believed to be responsible for restricting the formation
of new stars in quiescent galaxies.

Given the apparent paucity of mid-IR AGN in quiescent galaxies, the
contribution of these systems to the rapid growth of BHs must be
negligible in comparison to the AGN present in star-forming
galaxies. Hence, for all further analyses presented here, we neglect
the inclusion of quiescent galaxies in our spec-$z$ sample, as
identified using the UVJ diagnostic diagram, due to the systematic
lack of mid-IR AGN in these systems. Furthermore, by removing
relatively quiescent systems through our UVJ selection, our
morphological analyses that are designed to identify merging features
(described in Section~\ref{sec:imaging}) are not subject to
degeneracies arising from the existence of extremely long-lived
stellar shells that are readily identified in early-type systems
located within dense environments, and are unrelated to gas-rich
mergers.

\subsubsection{Separation of Obscured \& Unobscured AGN}

AGN identifications made at mid-IR wavelengths are relatively
independent of obscuration. Following simple AGN unification, Type-1
AGN are those where the accretion disk can be viewed almost directly,
with very little intervening gas or dust, while a Type-2 AGN is viewed
edge-on, and therefore has the disk emission and broad-line region
hidden from the line-of-sight by an optically thick torus surrounding
the central BH. However, as this torus isotropically reradiates the
AGN emission at IR wavelengths, a mid-IR AGN selection results in a
mixture of both Type-1 and Type-2 AGN. While the emission from both
these AGN populations dominate their SEDs at mid-IR, the
characteristic tail of the AGN accretion disk, which is typically
observed in the UV/optical, remains absent for only the Type-2
AGN. Hence, studies have revealed that a simple observed-frame
optical--IR color cut reliably separates unobscured Type-1 AGN from
their obscured counterparts (see Hickox et al. 2007; Hickox et
al. 2011; Chen et al. 2015).

\begin{figure}
  \begin{center}
    \includegraphics[width=\linewidth]{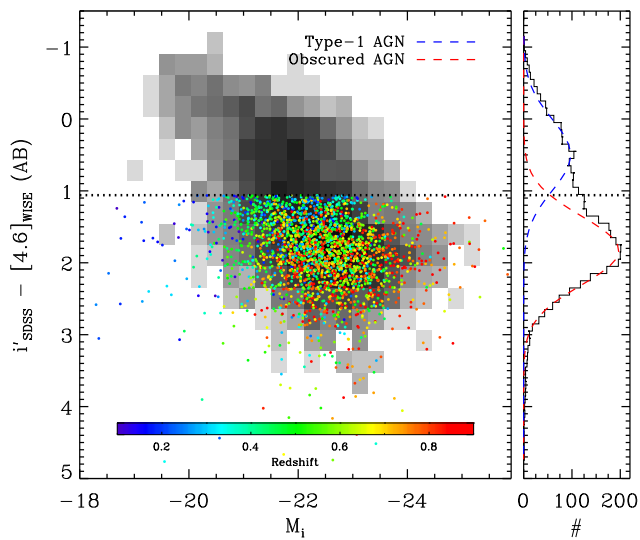} 
  \end{center}
  \caption{Observed optical--IR color versus absolute magnitude
    diagram used for separating our mid-IR selected AGN into obscured
    and unobscured subsamples (gray-scale contours). In $i$--[4.6]
    color, the AGN sample shows a distinct bimodality, with unobscured
    Type-1 AGN exhibiting bluer colors of $i-[4.6]<1.1$. Overlaid are
    the Type-2 AGN shown with rainbow colors to represent the source
    spectroscopic redshift. Right panel provides the histogram of the
    optical-IR color (black solid line) that is well characterized by
    the summation of two Gaussians, a type-1 AGN population (blue
    dashed) and an obscured AGN population (red dashed). The dotted
    lines is a simple cut that separates the AGN populations with
    minimum contamination.}\label{fig:type1type2}
\end{figure}

In Fig.~\ref{fig:type1type2} we present the distributions of our
mid-IR selected AGN sample in their observed-frame
$i_{\rm SDSS} - [4.6]_{\rm WISE}$ color. In a similar vein to
\cite{Hickox:2007aa}, we find that these optical--IR colors can be
characterized by two distinct Gaussian distributions, peaking at
$i_{\rm SDSS} - [4.6]_{\rm WISE} \sim 0.5$ and 1.9. Similar to Hickox
et al. (2007), we cut our AGN sample into obscured and unobscured
sub-samples using optical--IR color. We use a cut of
$i_{\rm SDSS} - [4.6]_{\rm WISE} = 1.1$, which is based on the
intersection of the Gaussian distributions. This serves to maximize
the number of AGN with the correct Type-1/2 classification, while
minimizing contaminants. We find 2,552 and 1,238 sources with
optical--IR colors that are red-ward and blue-ward, respectively, of
our $i_{\rm SDSS} - [4.6]_{\rm WISE} = 1.1$ cut. Inspection of the
SDSS spectroscopy for a subset of the AGN with
$i_{\rm SDSS} - [4.6]_{\rm WISE} \lesssim 1.1$ confirms the presence
of broad $H \beta$, $H \gamma$ emission lines and/or a strong blue
disk continuum.

We further show in Fig.~\ref{fig:type1type2} that the obscured and
unobscured AGN do not follow similar distributions when considered in
optical--IR color versus absolute $i$-band magnitude space. There is
an additional population of low redshift, low-luminosity
($M_i > -20.5$~mags), extremely blue objects with
$i_{\rm SDSS} - [4.6]_{\rm WISE} \lesssim -0.5$ that are not mirrored
in the obscured AGN population. These may be a set of AGN hosted in
very low-mass galaxies (e.g.,
\citealt{Satyapal:2014ad,Secrest:2015aa,Sartori:2015aa}) or a
population of low-metalicity blue dwarf galaxies
($M_* \lesssim 5 \times 10^{9} M_{\odot}$) with powerful young
starburst regions. These starbursts produce red colors in WISE that
are similar in practice to emission from AGN (e.g.,
\citealt{Hainline:2016aa}).

By contrast, there appears to be a population of luminous obscured AGN
at $M_i \lesssim -23.8$~mags that are not present in our Type-1 AGN
sample. At these luminosities, Type-1 AGN will most likely saturate
the HSC detector for relatively nearby systems and/or appear similar
to bright point sources at higher redshifts. These systems are
therefore preferentially removed from our sample during our initial
catalog selection by setting the {\tt is\_extended} flag. Such
dominance of the AGN over the host galaxy would hinder and bias our
determination of the host galaxy properties during the SED-fitting
process (e.g., stellar masses are known to be over-estimated for
Type-1 AGN) and during our morphological analysis presented in the
next section. Hence, to ensure the most unbiased measurements of the
AGN host galaxies, we select only the 2,552 obscured AGN with
$M_* > 5 \times 10^{9} M_{\odot}$ for all further analyses that
compare the AGN/galaxy properties.

\section{Identifying interacting and merging galaxies within HSC
  images}
\label{sec:imaging}

In this section we harness the exquisite sensitivity and spatial
resolution afforded to us by the HSC survey to provide a basic
morphological classification for each galaxy in our spec-$z$
sample. Using parametric and non-parametric metrics, combined with a
novel implementation of a Random Forest Machine Learning algorithm, we
separate our spec-$z$ galaxy sample into subsamples of major-mergers,
minor-mergers and irregulars, and non-interacting galaxies.

\subsection{Profile fitting with {\sc galfit}}

Image analysis techniques have been developed to produce parametric
measures that are capable of separating galaxies by their
morphological type. Using a-priori knowledge of a galaxy's structural
properties -- early-type galaxies have smooth, elliptical isophotes,
while late-type galaxies tend to be more disk-dominated with flatter
light-profiles -- it has been shown that even simple one or
two-dimensional decompositions of the light profiles are capable of
separating galaxies by their Hubble-type (e.g.,
\citealt{Kormendy:2009ab,Simard:2011aa}).

In order to analyze the size, morphology and stellar-light
distribution of the galaxies in our sample we begin by fitting a
single 2-dimensional Sersic profile \citep{sersic63} using {\sc
  galfit} \citep{Peng:2002aa} to the HSC $i$-band images. We extracted
100$\times$100~kpc postage stamps from the co-added data products
produced by {\tt hscPipe}, along with the associated variance image
and data mask. Point spread function (PSF) images are extracted from
the pipeline products on a source-by-source basis. Within {\tt
  hscPipe}, the PSF images are computed using the {\tt PSFEx} software
\citep{Bertin:2011aa} from 41$\times$41 pixel images of nearby stars to
determine the size and ellipticity of the PSF for each visit. These
PSFs are then co-added to replicate the average PSF of the co-added
image. The median PSF size for our sample is $\sim 0.6''$. See Bosch
et al. (2017) for further details on the computation of the PSF
images.

\begin{figure*}
  \begin{center}
    \includegraphics[width=\textwidth]{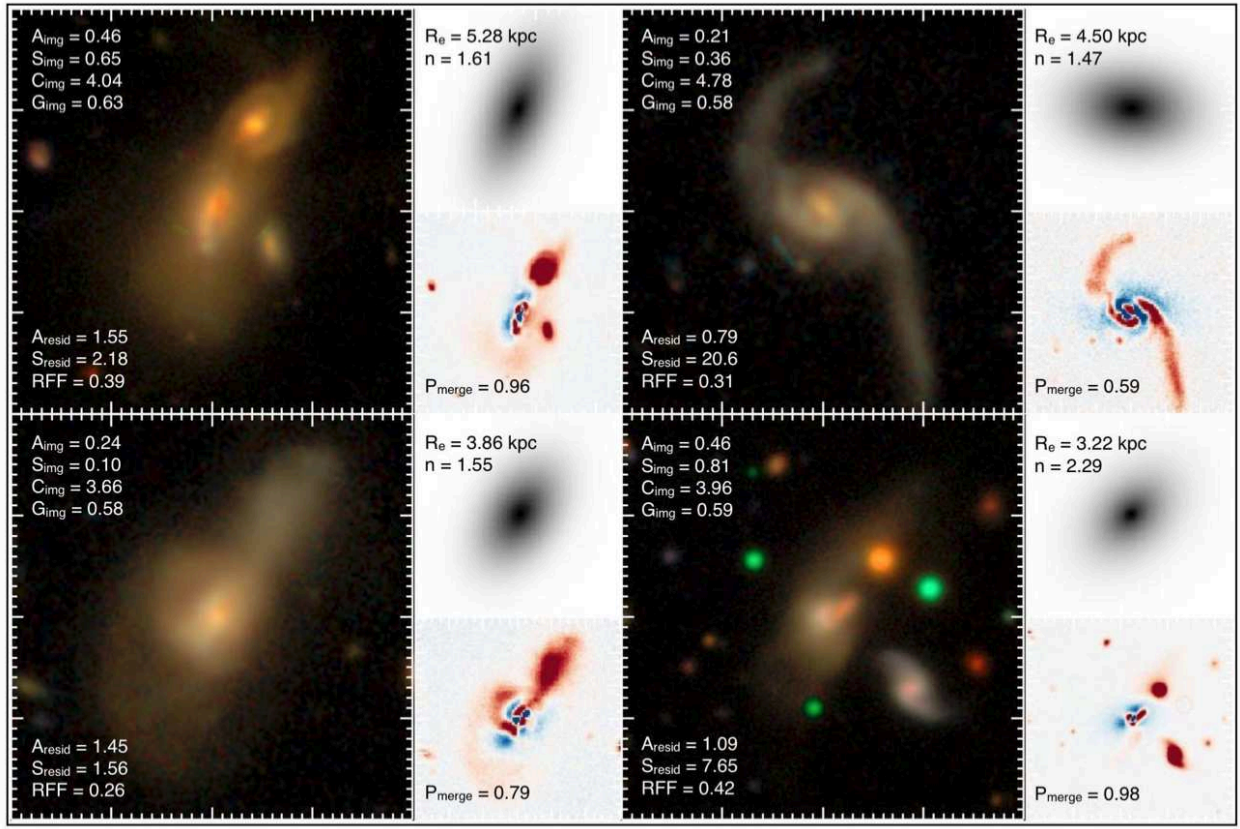} 
  \end{center}
  \caption{Four examples of the imaging analysis described in
    Section~\ref{sec:imaging} performed on our spec-z HSC galaxy
    sample. Large panels are K-corrected (pseudo-restframe) 3-color
    images; smaller inset panels are the best-fit Sersic model
    calculated using {\sc galfit} (upper) and the residual image ($i$
    -- model) with red color gradients for increasingly positive
    residuals and blue gradients for increasingly negative
    (lower). Labels provide the measures of asymmetry ($A_{\rm img}$),
    smoothness/clumpiness ($S_{\rm img}$), concentration index
    ($C_{\rm img}$) and Gini index ($G_{\rm img}$) calculated from the
    $i$-band image, as well the asymmetry ($A_{\rm resid}$),
    smoothness/clumpiness ($S_{\rm resid}$) and Residual Flux Fraction
    (RFF) calculated from the residual image. Interaction probability
    ($P_{\rm merge}$) determined from our implementation of a Random
    Forest Machine Learning algorithm in also given (see
    Section~\ref{sec:randomforest}).}\label{fig:galfit}
\end{figure*}

To measure a background level for each image, we used the full HSC
catalog, which is sensitive to sources with $i\sim 27$~mags, to
identify and mask all objects that lie within the postage stamp image
based on their catalog shape measurements and their {\tt Kron}
radii. We additionally applied the byte mask to those pixels
previously flagged by {\tt hscPipe} as erroneous. We fit a simple
2-dimensional linear profile to the non-masked pixels to assess any
overall background gradient within the image and determine a mean
background level in each pixel. We fill all areas within the
background image that were previously masked with Poisson noise
determined by the mean background level predicted for the individual
masked pixels. The measured background level was included as an input
to {\sc galfit}, and held fixed throughout the fitting procedure.

To create an input mask image for {\sc galfit}, we masked all sources
within the 100$\times$100~kpc postage stamp that had integrated
magnitudes at least 3~mag fainter than the target galaxy (i.e., a
factor 1:15 fainter in flux). All areas identified in the {\tt
  hscPipe} bad-pixel mask were also masked, and all bright point
sources were masked with shapes based on the ellipticities and radii
of the co-added PSF. For all remaining unmasked extended objects
within the image, we included an additional Sersic profile into the
{\sc galfit} fit centered at the position of the additional
galaxy. Hence, during the {\sc galfit} fitting procedure, will
simultaneously model all bright galaxies with the postage stamp
image. Our choice to mask objects determined to be at least a factor
$\sim 15$ fainter than the target objects allows us to simultaneously
model all components of possible major or minor mergers to at least
mass ratios of $1:10$ (i.e., allowing for variability in the
mass-to-light ratio).

We next extracted a sub-image of 50$\times$50~kpc centered around the
target galaxy along with the respective mask and variance images. This
sub-imaging approach has the advantage of limiting the computation
time with {\sc galfit}, while also maintaining that any large
(unrelated) sources, which may have significantly overlapping
isophotes with the region immediately surrounding the target galaxy
but may have centroids outside the sub-image, will still be
appropriately masked or have a Sersic profile assigned during the
fitting process. The source image, variance image and PSF model were
all used as inputs for {\sc galfit}.

In the upper-right sub-panels of Fig.~\ref{fig:galfit} we provide
examples of the best-fit 2-dimensional Sersic profile that were fit to
the target galaxy. For each galaxy, we extract the best fit parameters
for the Sersic profile, namely the Sersic index, $n$, and the
characteristic effective radius, $R_e$. $R_e$ is provided as a pixel
length within {\sc galfit}, which we convert to a physical scale in
kiloparsecs for all further analyses. In the next section, we use the
Sersic parameters and Sersic-profile subtracted images (residuals) to
compute metrics in order to identify interacting and non-interacting
galaxies.

\subsection{Automated Merger Detection using Supervised Machine Learning}
\label{sec:randomforest}

\subsubsection{Parametric \& non-parametric morphology indicators}

Many image analysis techniques have been developed to automatically
separate merging systems from non-interacting and/or isolated
galaxies, to varying degrees of success and accuracy. These methods
often make use of parameterizing the structures present in the image
of a given galaxy. In the previous section, we applied a 2-dimensional
Sersic profile to HSC postage stamp images, which was a simple
parametric approach for modeling the galaxy light distribution.

A tangential approach is to use non-parametric indices, which have
been developed to assess the distribution of light within an image in
order to separate/quantify a galaxy's Hubble class and/or interaction
stage (e.g.,
\citealt{Patton:2000aa,Patton:2002aa,Lin:2004aa,De-Propris:2007aa,Lin:2008aa,Robaina:2010aa,Bluck:2012aa,Glikman:2015aa}
for a recent review see \citealt{Conselice:2014aa}). Typical
non-parametric indices make use of the light concentration, asymmetry
and smoothness/clumpiness (hereafter, CAS measurements; see
\citealt{Bershady:2000aa,Conselice:2003aa}), as well as other measures
involving the Gini index and the second-order moments of the light
distributions (see \citealt{Abraham:2003aa,Lotz:2004aa,Lotz:2008aa}).

In the same spirit as these non-parametric indices, studies have now
begun to develop new metrics that implicitly incorporate parametric
measurements, resulting in hybrid parametric/non-parametric
indices. For example, the residual flux fraction (RFF;
\citealt{Hoyos:2011aa,Hoyos:2012aa}) measures the fluctuation of
counts in residual images of galaxies once a simple best-fit Sersic
profile has been subtracted. Residual images increase the contrast of
concentrated structures, as well as enhance low-surface brightness
features. Taken together, analysis of the residuals may better reveal
interaction signatures between galaxies that may otherwise be missed
in the original images.

Previous studies have determined that simple cuts on asymmetry and
smoothness ($A>0.35$ and $A>S$; \citealt{Conselice:2003aa}) or with
the Gini and $M_{20}$ parameters ($G > -0.14 \times M_{20} + 0.33$;
\citealt{Lotz:2004aa}) can produce a reliable ($\sim 50$\%) separation
of galaxies undergoing mergers in relatively nearby massive
systems. With the advent of new generations of telescopes and deep
surveys, like HSC, we are now able to resolve faint merger signatures
in large galaxy samples that were previously too weak to
identify. However, as sensitivity to low surface brightness material
increases, it becomes necessary to fine-tune our selection algorithms
to identify features of interest, particularly as long-lived tidal
debris, low surface brightness galaxies, and the outer parts of spiral
galaxies may all trigger the same indicators (e.g.,
\citealt{Greco:2017aa}).

Progress can be made by considering all of the information that can be
extracted from a combination of each of these different parametric and
non-parametric structure measures. Here we use a novel implementation
of a Machine Learning technique to provide a statistical measure of
the interaction state of a given system.

\begin{figure*}
  \begin{center}
    \includegraphics[height=21cm]{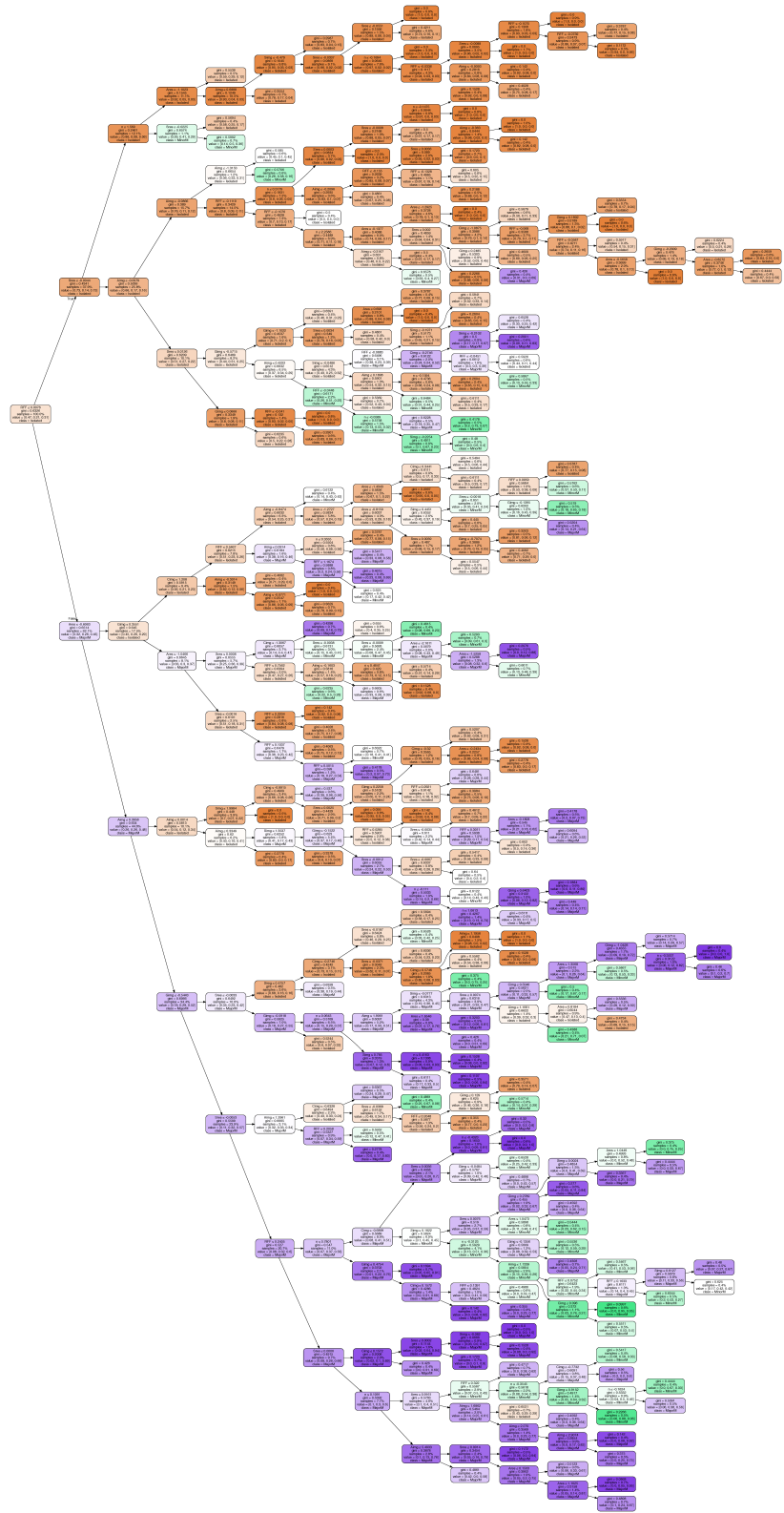} 
  \end{center}
  \caption{Example of a decision tree within our implementation of a
    Random Forest machine learning algorithm. The Random Forest is
    constructed from our representative sample of 5,900 visually
    classified galaxies. Each decision tree is formed from a bootstrap
    resampling of a subsample of 4,500 visually classified galaxies
    and is trained to identify objects based on three morphological
    classifications (1: non-interacting [orange]; 2:
    major/late-stage merger [purple]; 3: minor-merger/irregular
    [green]). Nodes are gradient color-coded depending on the purity
    of the classification decision (light colors have low purity, dark
    colors have high purity).}\label{fig:decision_tree}
\end{figure*}

As morphology ``features'' for our machine learning algorithm, we
measure the CAS parameters for each galaxy in our HSC spec-$z$ sample,
as well as the Gini and RFF indices. For the precise formulation of
these parameters we refer the reader to Section 2.3 of the review by
\cite{Conselice:2014aa} and \cite{Hoyos:2012aa}. We measure each of
these indices on the 50$\times$50~kpc $i$-band postage stamp galaxy
images. Following \cite{Hoyos:2012aa}, we also compute the asymmetry
and smoothness/clumpiness parameters on the residual flux images
(i.e., the $i$-band image after subtraction of the best-fit Sersic
model for the galaxy determined following the method outlined in the
previous section). These non-parametric indices are combined with the
parametric measurements of the best-fit Sersic profiles to provide a
suite of morphological parameters (hereafter, `features') that we use
to determine the interaction state of the galaxy through `automated
classification'.

\subsubsection{Training a Random Forest Classifier}

The goal of automated classification frameworks is to determine a
model that describes some in-hand data for a set of objects whose
`science classification' is known a-priori. This model is then applied
a new set of objects, whose classifications are unknown, and then used
to predict a class or probability of a given classification for each
new object. Several forms of data-driven automated classification
schemes have been used to solve an abundance of astrophysical
problems, such as Gaussian mixture models, Bayesian networks, neural
networks, and support vector machines
\citep[e.g.,][]{Goldstein:2015aa,Moolekamp:2015aa,Williams:2016aa,Melchior:2016aa,Avestruz:2017aa}. A
conceptually simple, extremely efficient, and yet powerful
classification method, which is becoming popular throughout astronomy,
is that of decision-tree learning.

Decision trees are supervised non-parametric classifiers that remain
efficient even when attempting to capture complicated feature-based
structures. They naturally handle multiple classification schemes, and
are relatively robust to outliers. However, tree models tend to have
high variance. Due to the hierarchical structure of the trees, even
small changes in the top levels of a training tree, induced by random
selection of the variables used to split nodes, can produce vastly
different trees on subsequent nodes. Also, while large trees will, by
design, always fit the training data very well, a specific large tree
may not generalize well to test data. This process is akin to
over-fitting in simple regression.

\begin{figure*}
  \begin{center}
    \includegraphics[width=\linewidth]{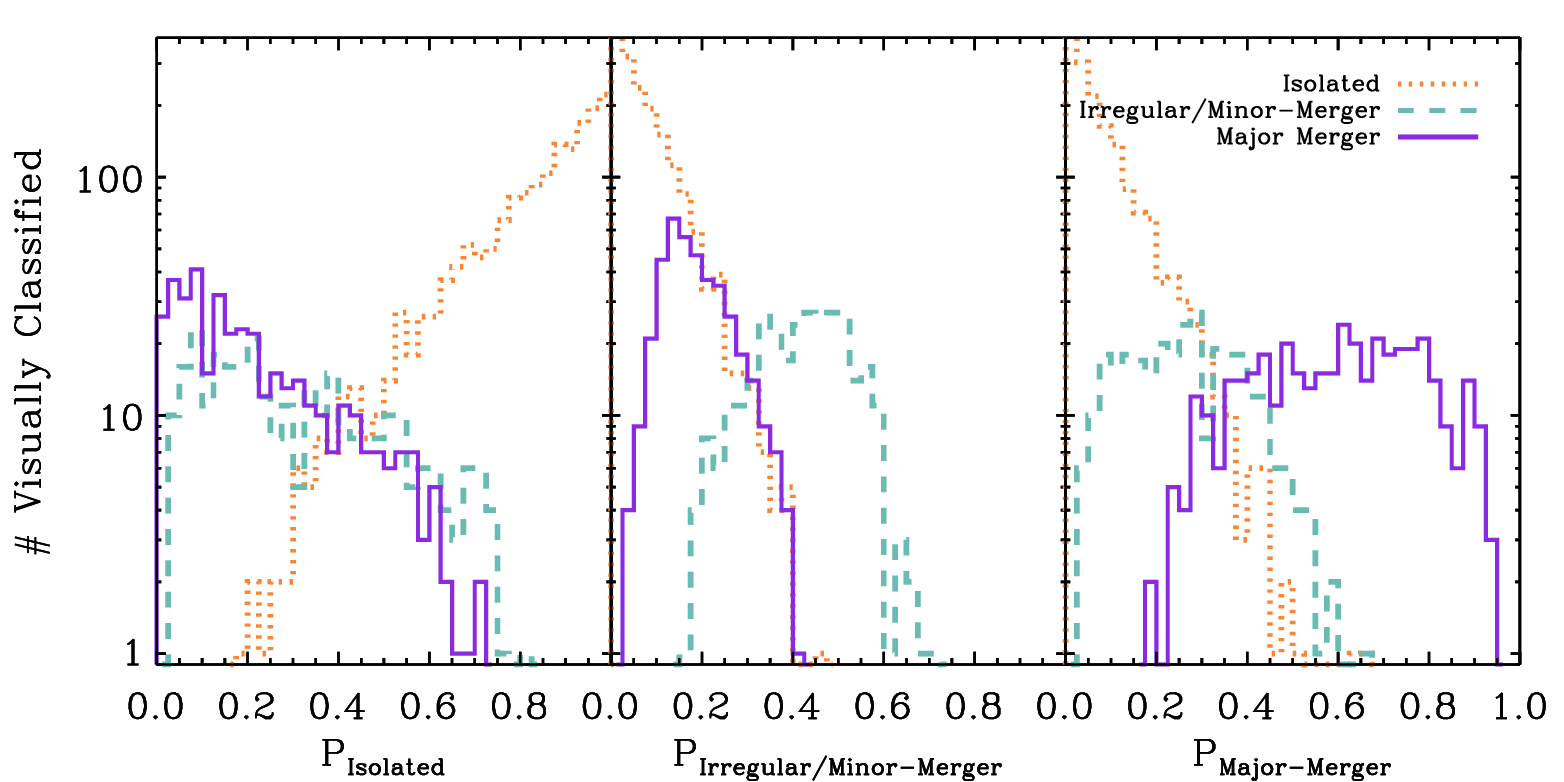} 
  \end{center}
  \caption{Distributions of probabilistic merger-state classifications
    assigned by our implementation of Random Forest Machine Learning
    algorithm to 1,400 visually classified galaxies. The Random Forest
    was trained on an independent sample of 4,500 visually classified
    galaxies randomly selected from our main HSC spec-z sample. Left:
    probability of being an isolated galaxy ($P_{\rm Isolated}$);
    Center: probability of being an irregular galaxy or a minor-merger
    ($P_{\rm irregular/minor-merger}$); Right: probability of being a
    major merger ($P_{\rm Major-Merger}$). Distributions are split as
    a function of their visual classification (isolated;
    irregular/minor-merger; major merger), see
    Section~\ref{sec:randomforest} for further
    details.}\label{fig:allprobs}
\end{figure*}

Noise in the final classifications can be reduced by considering
multiple decision trees for a given dataset, so-called `Random
Forests'. Random Forest classifiers fit multiple decision trees to
bootstrap subsamples of a given training set. The final classification
for an object is then the average of the classifications produced by
the individual bootstrap decision trees, which naturally provides a
(pseudo-)probability for the classification (driven by the input training
data) while controlling for over-fitting of the data.

\begin{table}
  \caption{{\tt RandomForestClassifier} Initiation Parameters}
  \begin{center}
  \begin{tabular}{lc}
      \hline \hline
      \multicolumn{1}{c}{Parameter} & 
      \multicolumn{1}{c}{Value} \\ 
      \hline
      {\tt input features} & $C_{\rm img}$; $A_{\rm img}$; $S_{\rm img}$; $G_{\rm img}$; \\
                           & $A_{\rm res}$; $S_{\rm res}$; RFF; $R_e$; $n$ \\
      {\tt n\_estimators}  & 1000 \\
      {\tt criterion}      & {\it gini} \\
      {\tt max\_features}  & $\sqrt{9}$ \\
      {\tt max\_depth}     & 15 \\
      {\tt min\_samples\_split} & 12 \\
      {\tt bootstrap}      & True \\
      {\tt warm\_start}    & False \\
      {\tt class\_weight}  & {\it balanced} \\
      \hline \hline
    \end{tabular}\label{tab:randforest}
    \end{center}
% \begin{tabnote}
% This is table note.
% \end{tabnote}
\end{table}

To build our training sample, we visually classified the
50$\times$50~kpc k-corrected 3-band HSC images for a random sample of
5,900 galaxies in our spec-$z$ sample that were deemed to be
star-forming based on their position in the UVJ diagram. The specific
visual classification scheme involved the identification of (1)
irregular/disturbed/torqued morphologies, (2) double-nuclei/late-stage
merger, (3) evidence for interaction with a distinct companion galaxy,
(4) regular morphologies with no evidence for recent interaction, or
(5) too-small to conclusively identify. To normalize the responses of
the seven expert classifiers, we averaged the individual visual
classifications for a test subsample of 600 galaxies, and then
weighted the responses accordingly for the remaining visual
classifications. For galaxies that were clearly undergoing or had
recently undergone an interaction, our expert classifiers were in
strong agreement that at least one of the interaction classifications
was valid. However, we noted significant variance among the experts
when attempting to separate these different signatures of
galaxy--galaxy interactions. As such, we elected to consolidate our
visual classifications for interacting systems, as we determined that
this provided a cleaner separation between interacting and
non-interacting galaxies.\footnote{For our visually classified
  training sample we split the galaxies determined to be ``(3)
  evidence for interactions with a companion'' by the flux ratio of
  the two interacting systems. In accordance with previous studies, we
  used a demarcation of 1:4 in flux ratio to denote major and minor
  mergers. Interacting galaxies with ``(2) double-nuclei/late-stage
  mergers'' were additionally considered major-mergers, while
  irregular morphologies were considered to be minor-mergers.}

Examples of four systems determined to be major-mergers from their
visual classifications are shown in Fig.~\ref{fig:galfit}. Each of
these systems are clearly at different stages of merging. In terms of
the interaction classification outlined by \cite{Veilleux:2002aa},
these galaxies would be classified as IIIa:{\it wide binary} (right
column), IIIb:{\it close binary} (top-left) and IV:{\it Merger}
(bottom-left).\footnote{We note that sources with the interaction
  classification of `I:first approach' would be considered as
  `non-interacting' in our visual classifications as the galaxy disks
  have not yet been perturbed.} These four examples exhibit relatively
wide ranges in parameters such as their smoothness/clumpiness
$\sim 0.1-0.8$ (typical values in the range -0.5--1.5), but have
narrow ranges in Gini ($\sim 0.6$) and RFF ($\sim 0.3-0.4$). The role
of our Random Forest implementation will be to search for correlations
between the visual classifications and the specific values/ranges of
these features.

Our visually-classified training sample was split to provide an input
of 4,500 galaxies used to construct the decision trees, and an
independent subsample of 1,400 galaxies to test the output
classifications of the Random-Forest classifier. We used the publicly
available Python-based {\tt RandomForestClassifier} code provided as
part of the {\tt scikit-learn} package \citep{scikit-learn} to build
the decision trees. The input features for the decision tree
construction (see Table\ref{tab:randforest}) were the concentration
($C_{\rm img}$), asymmetry ($A_{\rm img}$), smoothness/clumpiness
($S_{\rm img}$) and Gini ($G_{\rm img}$) indices measured from the HSC
$i$-band images, the RFF, asymmetry ($A_{\rm res}$) and
smoothness/clumpiness ($S_{\rm res}$) indices measured from the
residual (galaxy--Sersic model) image, and the Sersic index and $R_e$
measured from the best-fit model.

To avoid importance bias of a particular input feature, we first
normalize the distributions of each feature to have mean zero and
unity variance before inputting to the Random Forest generator. The
Random Forest is initiated with the parameters shown in Table 1, and
then trained to identify galaxies based upon the three morphological
classifications assigned during our visual classifications: 1:
non-interacting (inclusive of Stage I pre-mergers); 2: major-merger
(inclusive of Stage II--IV mergers); 3: minor-mergers (inclusive of
Stage V irregulars). The final assigned classification is then the
average of the `votes' from each of the 1000 decision trees, i.e., the
fraction of trees that assign a classification of `isolated' is
$P_{\rm isolated}$.

An example of one of the 1000 decision trees in the Random Forest is
shown in Fig.~\ref{fig:decision_tree}. After experimentation, the
branches are pruned to not allow depths beyond 15 nodes, though most
branches terminate before this as we set a minimum threshold of $>12$
sources for a new node to be created. In the example presented in
Fig.~\ref{fig:decision_tree} we find that in the initial node of the
tree (left-most box in the diagram), that a relatively neutral cut in
RFF (0.0076 in normalized units) ultimately results in a strong
overall distinction between interacting and non-interacting
galaxies. All subsequent nodes leading upwards and away from the
initial node (i.e., training objects with RFF$\leq 0.0076$) are, in
general, colored orange, denoting non-interacting galaxies. By
contrast, subsequent nodes leading downwards and away from the initial
node (i.e., training objects with RFF$> 0.0076$) are more likely to
result in nodes containing interacting galaxies (colored either
purple:major-merger or green:irregular/minor-merger).

Furthermore, in Fig.~\ref{fig:decision_tree} we show that
minor-mergers are difficult to distinguish from major mergers and
non-interacting galaxies. The minimum node value (i.e., the number of
connecting nodes required to reach a node from the initial [left-most]
node) of an irregular/minor-merger classification is 4, with the
majority of the irregular/minor-merger leaves not being identified
until node $>7$. From a decision tree stand-point, minor-mergers then
become a sub-category of the more dominant isolated and major-merger
classifications, making their robust identification complex.

Using our training visual classification sample, we additionally
calculated the importance of the input features that went into
producing our Random Forest classifier. The importance can be thought
of as the fraction of useful information that is used by the
classifier during the construction of a decision tree, with the sum of
importances, $I$, over all features equaling unity. The most important
features, averaged over all trees, were $S_{\rm res}$, $A_{\rm img}$
and RFF, each with $I \sim 0.17-0.21$, while the $C_{\rm img}$ was the
least useful with $I \sim 0.03$.

\subsubsection{Testing the Random Forest Classifier}

\begin{figure}
  \begin{center}
    \includegraphics[width=\linewidth]{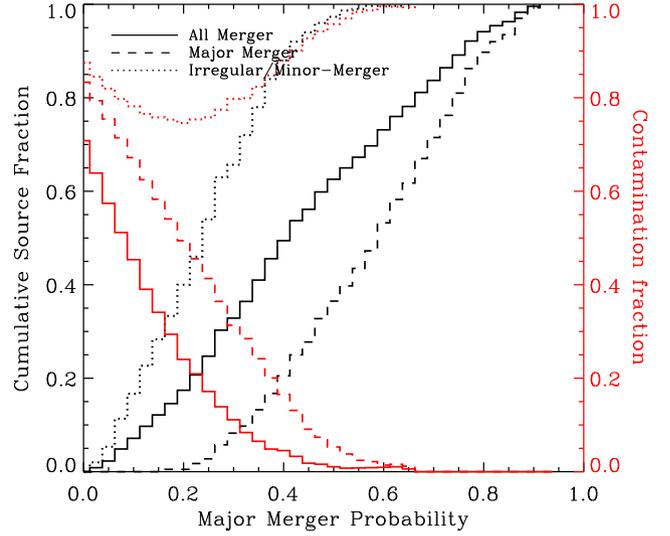} 
  \end{center}
  \caption{Fraction of sources in our test set of 1,400 visually
    classified galaxies as a function of the probability of a
    particular system being a merger. Merger probabilities are
    computed during the implementation of a python-based Random Forest
    Machine Learning algorithm trained on an independent set of 4,500
    visually classified galaxies in our main HSC spec-z
    sample. Dashed, dotted and solid lines are those galaxies visually
    classified to be major mergers (flux-ratio $>1:4$), minor mergers
    $1:4-10$, and major or minor mergers, respectively. Red lines
    provide the fraction of objects with a given $P_{\rm merge}$ that
    are determined to not be the given merger classification (i.e.,
    for major mergers, contaminant populations are non-interacting
    galaxies and minor-mergers).}\label{fig:prob_merge}
\end{figure}

To assess the ability of our Random Forest for providing reliable
probabilistic classifications to the remainder of our HSC spec-z
galaxy sample, we applied the trained Random Forest to our `test
sample' of 1,400 visually classified galaxies that were not used
during the training of the classifier. In Fig.~\ref{fig:allprobs} we
provide the distributions of the classification probabilities for our
test sample separated by their visual classifications. For each of
classification probability, the distribution of the true visual
classified objects peak at higher probability values. Indeed, it is
clear from the $P_{\rm isolated}$ histograms that we can cleanly
recover a sample of isolated galaxies with a cut of
$P_{\rm isolated} > 0.7$, with little or no contamination from
interacting galaxies. However, this of course does not recover the
full population of isolated systems, as this population of objects
begins to mix significantly with objects towards lower values of
$P_{\rm isolated}$. This is also mirrored in the distributions of
$P_{\rm minor-merger}$ and $P_{\rm major-merger}$. 

In Fig.~\ref{fig:prob_merge} we further explore contamination to a
major merger sample when applying a threshold in $P_{\rm merger}$. As
expected, we find galaxies assigned to have high values of
$P_{\rm merger}$ by the Random Forest are increasingly more likely to
actually be major mergers based on their visual classification. We
find that while a threshold of $P_{\rm merger} > 0.33$ would provide a
sample that is $\sim 90\%$ complete towards major mergers, $\sim 30$\%
of the sample would be contaminated by non-interacting galaxies and
minor-mergers/irregulars. Based on our Random Forest and limited
training sample, we cannot yield a truly pure sample of major mergers
that is more than $\sim 39\%$ complete. However, in the range
$0.32 < P_{\rm merger} < 0.67$ it is clear we suffer from only mild
contamination ($\sim 10\%$), and only $\sim 1/3$ of the sample
contamination arises from isolated galaxies. Hence, a cut of
$P_{\rm merger} > 0.32$ yields a relatively clean sample of
interacting systems (i.e., minor$+$major merger), while a cut of
$P_{\rm merger} > 0.46$ yields a sample of major-mergers that is over
$> 75\%$ complete and suffers less than $\sim$10\% contamination, the
majority of which arises due to minor-mergers and irregulars, which
themselves may have somewhat ambiguous visual classifications given
the almost arbitrary demarcations that are made between the visual
classes. Finally, we also tested for any effect to the classifications
due to the presence of an unobscured AGN, which may have been
incorrectly classified as a Type-2 AGN from our IR--optical
color-cut. While we do include a PSF model during our {\sc galfit}
analysis, we found that the presence of a Type-1 AGN still marginally
steepens the Sersic index, significantly increases the concentration
index, and lowers the asymmetry value. These are each due to the
galaxy light being partially contaminated by the AGN. Irrespective of
whether the source was visually classified as an non-interacting or
major-merger, we found this typically lowered the value of
$P_{\rm merger}$, resulting in the source being more likely to be
classified as a non-interacting galaxy. As such, we note here that the
presence of Type-1 AGN in our sample may artificially increase the
merger fraction in non-interacting galaxies, and hence these will
dilute the signal from AGN being intrinsically preferentially hosted
in major mergers in the next section. In the next section, we use
these automatic classification probabilities to construct relatively
robust samples of non-interacting isolated galaxies, major mergers,
and a set of interacting (irregular$+$minor-merger$+$major-merger)
galaxies, and investigate the incidence of AGN in these systems.

\section{Results}
\label{sec:results}

Despite the theoretical successes of BH--galaxy co-evolution models to
explain observed present-day galaxy populations, observational
evidence for the presence of an on-going merger and the concurrent
rapid growth of BHs, which is now a required ingredient of galaxy
formation simulations, remain elusive. In this section we use the
morphological/interaction probabilities derived using our
implementation of a Random Forest machine learning algorithm to assess
the incidence of AGN in carefully constructed
statistically-significant samples of major mergers, minor mergers and
irregulars, and non-interacting galaxies.

\subsection{Incidence of AGN in interacting and non-interacting
  galaxies}
\label{sec:incidence}

\begin{figure*}
  \begin{center}
    \includegraphics[width=0.95\textwidth]{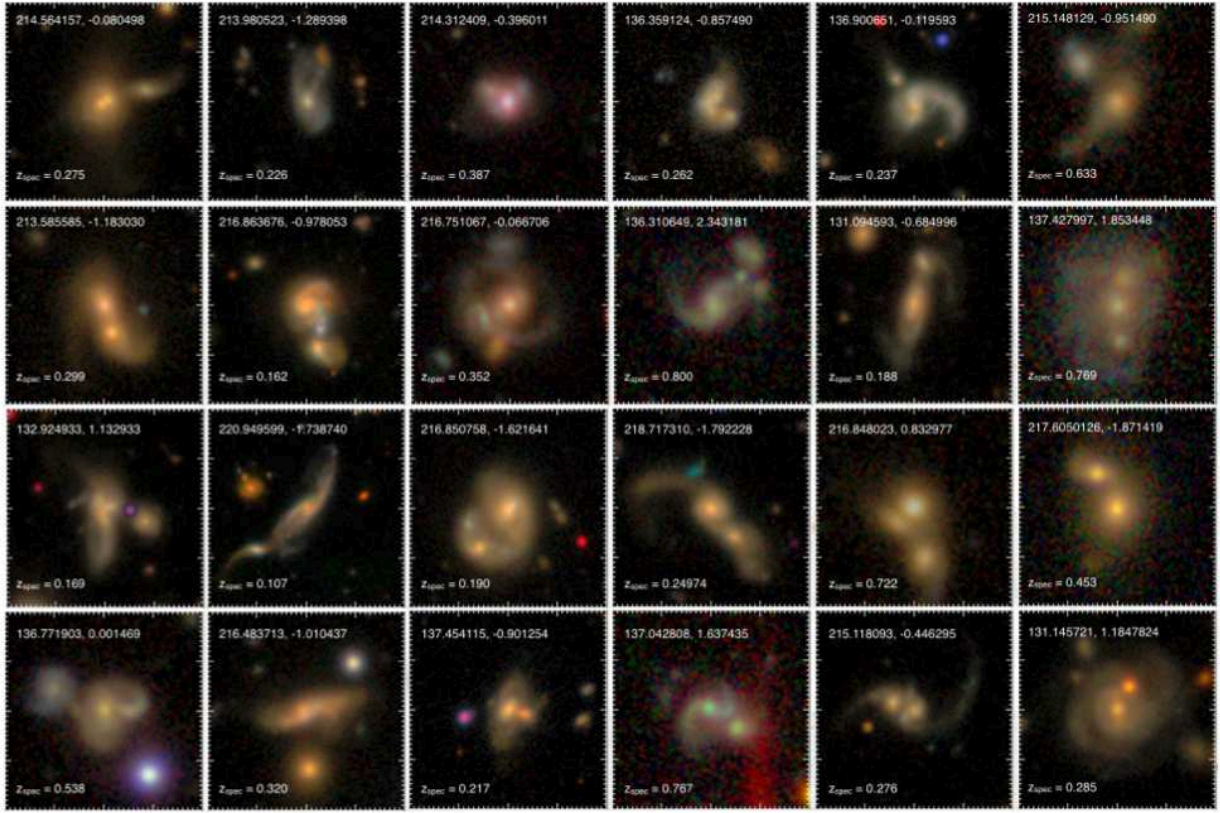} 
  \end{center}
  \caption{Examples of K-corrected 3-color 50$\times$50~kpc HSC images
    of the 4,449 gas-rich major-merger candidates at $0.1 < z < 0.9$
    selected to have $P_{\rm major-merger} > 0.46$ based on our Random
    Forest Classifier described in
    Section~\ref{sec:randomforest}. These merger candidates cover a
    wide-range in interaction state, stellar mass ratio and
    redshift.}\label{fig:merger_eg}
\end{figure*}

AGN activity is a highly stochastic process, with changes in accretion
rate that typically occurs on time-scales that are much shorter than
longer lived galactic processes, such as changes in stellar mass,
star-formation rate, or even merger-stage. Thus, it is more robust to
probe the average AGN property (i.e., averaging over BH accretion
variability) as a function of the longer timescale galactic process
(see \citealt{Hickox:2014aa}). Hence, we now investigate the average
incidence of AGN based on host-galaxy interaction stage by harnessing
three galaxy samples: 1) major mergers; 2) all interacting galaxies
(including major merger, minor mergers, and irregular systems), and 3)
non-interacting galaxies.

\subsubsection{Selecting galaxies \& AGN in bins of interaction type}

In Section~\ref{sec:randomforest} we determined that a threshold of
$P_{\rm major-merger} > 0.46$ in our trained Random Forest classifier
recovers $\sim 75\%$ of the major-mergers present in our
visually-classified test sample. A sample of major-mergers defined by
such a cut suffers contamination at the $\sim 7$\% level from
minor-mergers and irregular galaxies, and $< 3$\% from non-interacting
galaxies. Applying this threshold in $P_{\rm major-merger}$ provides a
clean sample of 4,449 gas-rich major mergers at $0.1 < z < 0.9$. In
Fig.~\ref{fig:merger_eg}, we show a random set of examples of major
mergers identified by our Random Forest Classifier, which reside in a
subregion of the HSC survey. The diversity of the sample of major
mergers in their interaction state, mass ratio, number of systems, and
redshift is clearly apparent. This is mainly driven by our large
training sample of 4,500 visually classified systems covering wide
ranges in galaxy properties (including interaction state), as well as
the sensitive HSC imaging that is capable of detecting the low surface
brightness emission associated with merging, which may be otherwise
missed in shallower wide field surveys.

We define two additional galaxy samples: (1) a set of galaxies at all
stages of interaction, which include major mergers, minor mergers and
irregular systems; and (2), a control set of non-interacting/isolated
galaxies. Following our testing in the previous section, we invoke a
threshold of $P_{\rm merger} > 0.32$ {\tt or}
$P_{\rm minor-merger} > 0.40$ to define the set of interacting
galaxies, which yields an interacting sample of 5,594 systems. The
non-interacting star-forming galaxies are defined by
$P_{\rm isolated} > 0.7$, which provides a sample of 12,513 galaxies,
with $\ll 1$\% contamination from interacting systems.

\begin{figure*}
  \begin{center}
    \includegraphics[width=0.85\textwidth]{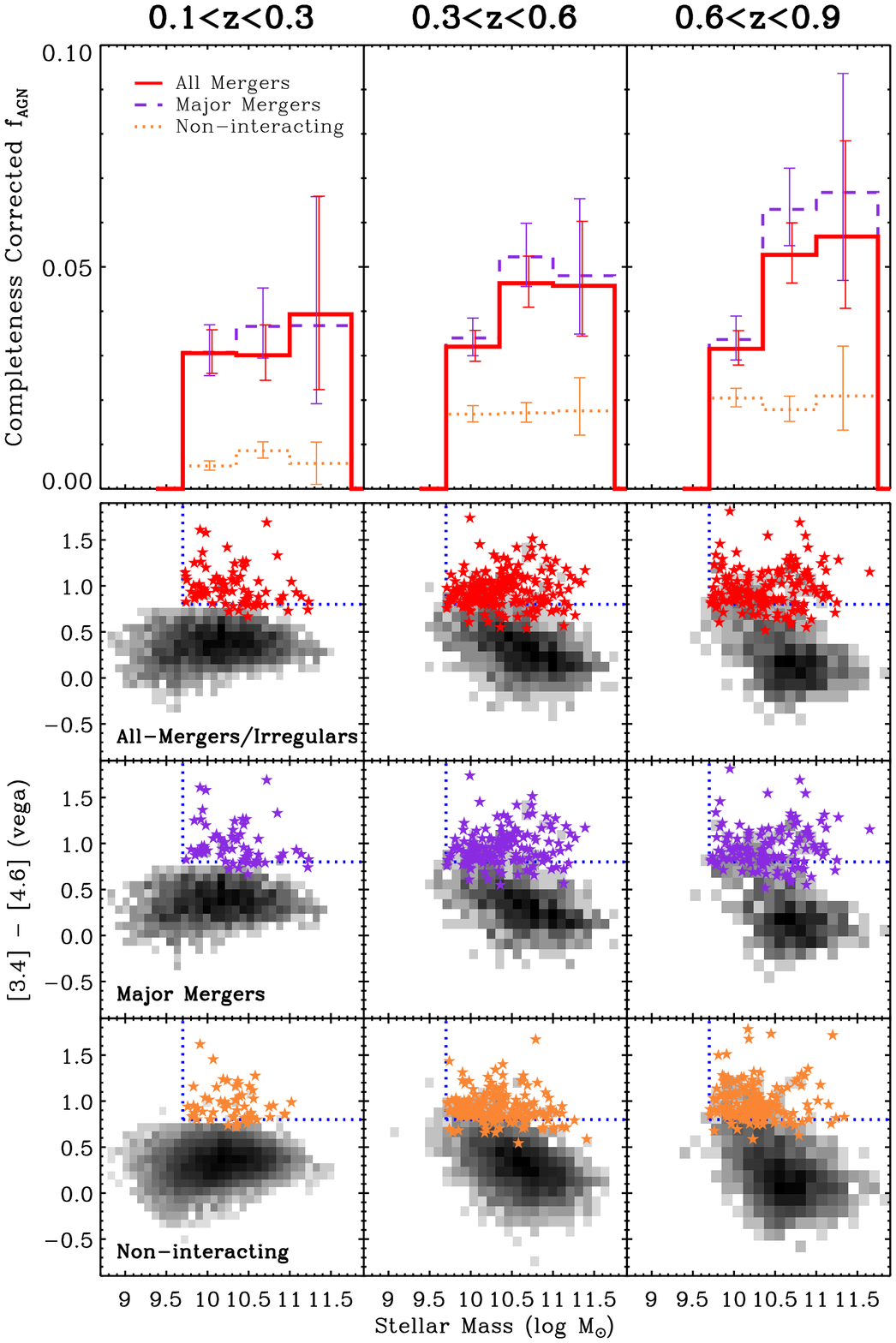} 
  \end{center}
  \vspace{-0.3cm}
  \caption{{\bf Lower Panels:} WISE [3.4]--[4.6] color (vega) as a
    function of stellar mass (in logarithmic units of $M_{\odot}$) for
    three morphologically selected samples -- non-interacting galaxies
    ($P_{\rm isolated} > 0.7$; orange; bottom row), major mergers
    ($P_{\rm major-merger} > 0.46$; purple; center row), and all
    interacting galaxies (major-merger$+$minor-merger$+$irregulars;
    $P_{\rm major-merger} > 0.32 \bigvee P_{\rm minor-merger} > 0.4$;
    red; top row). For merging/interacting systems, reported stellar
    masses are those of the most massive galaxy in the pair. Panels
    columns are split by redshift, using the same cuts as in
    Fig.~\ref{fig:uvj}. Galaxies identified to host IR-luminous Type 2
    AGN using the diagnostics presented in
    Figs.~\ref{fig:wise_color_color} and \ref{fig:type1type2} are
    shown with star symbols and colors denoting morphological
    classification. Dotted lines illustrate the AGN selection
    boundaries for galaxies with $M_* \gtrsim 5 \times 10^9 M_{\odot}$
    and [3.4]--[4.6] $>$ 0.8 \citep{Stern:2012aa} {\bf Top Panels:}
    Histograms of AGN fraction as a function of stellar mass
    constructed in redshift bins matched to the lower panel. AGN
    fractions in a given redshift bin are separated by morphological
    classification -- non-interacting galaxies (orange dots); major
    mergers (purple dashed); all interacting galaxies (red solid). The
    AGN fractions are also corrected for incompleteness of non-WISE
    detections in the two highest redshift bins. We find in each
    redshift bin that galaxies undergoing mergers are a factor
    $\sim 2-3$ more likely to contain AGN than non-interacting
    galaxies, and this is independent of stellar
    mass.}\label{fig:money}
\end{figure*}

In Fig.~\ref{fig:money} we show the distributions of the three samples
in their WISE [3.4]--[4.6] color and stellar mass, separated by three
redshift bins ($0.1<z<0.3$, $0.3<z<0.6$, and $0.6<z<0.9$). The
[3.4]--[4.6] color is indicative of the dust temperature, and incident
SF or AGN activity. Additionally, we highlight each galaxy that has a
significant AGN contribution to the mid-IR continua, and thus has a
relatively high [3.4]--[4.6] color ($\gtrsim 0.7$). The AGN were
previously identified based on their 2 or 3 band mid-IR colors, as
measured from their WISE photometry (see Section~\ref{sec:agnID});
crucially, the AGN selection was performed independently of the
morphology and interaction state of the galaxy. 

Previous observations have suggested that the AGN fraction (above some
particular threshold in AGN luminosity) rises steeply as a function of
stellar mass (e.g.,
\citealt{Xue:2010aa,Aird:2012aa,Bongiorno:2012aa,Mullaney:2012aa}). Thus,
it is important to consider stellar mass matched samples when
assessing the relative incidence of AGN in our
morphological/interaction-state samples. Here we provide the quoted
mass-completeness limits towards star-forming galaxies of the
spectroscopic surveys, from which our main spec-$z$ sample derives. As
shown previously in Fig.~\ref{fig:fig1}, in the lowest redshift bin,
our sample is dominated by galaxies identified in the GAMA-DR2 survey,
which is complete to $M_* \sim 5\times 10^9 M_{\odot}$ for
star-forming galaxies with restframe $g-i < 0.5$ \citep{Taylor:2011aa}
at the median redshift of the bin ($z\sim0.2$). Furthermore, at
$z\sim0.5$, our sample is mainly comprised ($\sim 75$\%) of galaxies
drawn the PRIMUS and VIPERS surveys, which for star-forming galaxies
at $z \sim 0.5$ are complete to $M_* \sim 2\times 10^9 M_{\odot}$
\citep{moustakas13} and $M_* \sim 3\times 10^9 M_{\odot}$
\citep{Davidzon:2013aa}, respectively. We note that while these surveys claim
completenesses toward star-forming galaxies of
$M_* \sim 5\times 10^9 M_{\odot}$, we ultimately do not require our
sample to be complete in stellar mass, as the AGN fractions we present
in the proceeding sections are compared in relatively between
interacting and non-interacting galaxies.

\subsubsection{WISE Completeness Corrections}

In each of our considered redshift bins, our star-forming parent
sample is roughly mass complete at around
$M_* \sim 5\times 10^9 M_{\odot}$. However, as we advance in redshift,
we systematically miss low-mass galaxies in our sample due to the flux
limit of the WISE survey. This is observed in Fig.~\ref{fig:money} as
a deficit of sources in the $0.3<z<0.6$ and $0.6<z<0.9$ bins that have
blue [3.4]--[4.6] colors and
$M_* \lesssim 2 \times 10^{10} M_{\odot}$.  This subsection outlines
our statistical corrections for this incompleteness brought about by
the WISE--HSC cross-match.

By comparison of the fractions of WISE detected and non-detected
galaxies at $0.3<z<0.6$, we find that for galaxies in our spec-$z$
sample with $M_* \sim 5 \times 10^{9} M_{\odot}$, $\sim 85$\% are not
detected by WISE. Conversely, at the same redshift, only $\sim$5\% of
the galaxies with $M_* \sim 10^{11} M_{\odot}$ are not detected in
WISE. If we compare these fractions of WISE non-detections to our
lowest redshift bin ($0.1<z<0.3$), we find that only $\sim$3\% and
$\sim$2\%, respectively, of the galaxies are not detected in WISE in
the same mass bins. This demonstrates that we are essentially complete
towards WISE detections at low redshift, and that the loss of low-mass
blue WISE color sources as a function of redshift can be almost
entirely attributed to the flux limit in WISE.

From Fig.~\ref{fig:money} we can deduce that the loss of
low-mass blue WISE color sources is relatively independent of
morphology classification. Two-sample K-S tests show that there is no
significant evidence for a difference between $M_*$ distributions of
the non-interacting, major merger and all-merger samples within a
given redshift bin.

IR-luminous AGN are selected to be more luminous and redder in WISE
than non-active galaxies. As such, incompleteness may not affect AGN
in the same way as non-active galaxies. To test this, we can compare
the raw AGN fraction (i.e., no corrections for underlying
completeness) between two redshift and stellar mass bins. For galaxies
with $M_* \sim 10^{11} M_{\odot}$ (a stellar-mass bin which is
unaffected by WISE-completeness issues) the AGN fractions at
$0.1<z<0.3$ and $0.3<z<0.6$ are consistent with each other:
$\sim 2.4$\% and $\sim 2.3$\%, respectively, and we therefore see no
evolution in AGN fraction between the redshifts in this mass
bin. However, for galaxies with $M_* \sim 5 \times 10^{9} M_{\odot}$,
the AGN fractions increase by almost an order of magnitude from
$\sim 1.9$\% at $0.1<z<0.3$ to $17.2$\% at $0.3<z<0.6$. Such a large
jump in AGN fraction in a small redshift range is wholly unphysical,
and strongly suggests that while we are not detecting non-active
galaxies in WISE for galaxies with
$M_* \sim 5 \times 10^{9} M_{\odot}$, we do still identify those
galaxies containing AGN. Thus, in order to calculate the AGN fractions
presented in the following sections, we need only statistically
account for the non-active WISE undetected objects in each $M_*$ bin
as a function of $z$. Indeed, we show in the next section that by
making the assumption that we are HSC--WISE cross-match only misses
non-active galaxies, and correcting for this incompleteness, the AGN
fractions are found to be similar in each mass bin across the full
redshift range considered.

Based on our parent sample, in each redshift bin we compute the $M_*$
distributions of the galaxies that are not detected in WISE during our
WISE--HSC cross-match. Under the informed assumption that none of
these systems contain AGN, we use the ratio of the $M_*$-distributions
between the WISE detected and non-detected to systems to normalize the
AGN fractions presented in the next subsection.\footnote{For
  consistency, we combine all galaxies within a redshift bin (i.e.,
  independent of interaction state), and compute a single correction
  function. However, we note that calculating completeness corrections
  that are interaction-state specific, and applying these to our AGN
  fractions presented in Fig.~\ref{fig:money} has no overall effect on
  our conclusions. This is fully expected given that the normalized
  two-dimensional distributions in $M_*$--WISE-color space are similar
  between the morphological samples in a given redshift bin.} As our
lowest redshift bin is relatively complete towards WISE detections,
our computed corrections in this redshift bin are factors of
$\sim 1.02$--1.03. However, these corrections become large,
$\gtrsim 10$, in the lowest mass bins at higher redshift. Ultimately,
these completeness corrections allow us to qualitatively compare AGN
fractions across the redshift bins. But crucially, as the corrections
are applied irrespective of the interaction classification, they do
not effect our conclusions when comparing fractional differences
between AGN fractions at fixed $M_*$ and $z$.

\subsubsection{AGN fractions at fixed $M_*$ and $z$}

For a given redshift bin, we construct 3 equal stellar mass bins with
width 0.7 dex (a factor $\sim 5$) for each of our morphological
samples. Accurate photometric measurements, and hence, stellar mass
measurements, for merging galaxies are non-trivial for the most
distant sources in our sample. As stated previously (see
Section~\ref{sec:speczsample}), based on our comparison between HSC
and Hubble Space Telescope data in the COSMOS field, we determined
that the photometric measurements for distant mergers to be accurate
to $\pm 0.3$~mags, resulting in a factor $\sim 1.3-1.5$ uncertainty
in derived $M_*$ (dependent on typical mass-to-light ratios). As such,
we conservatively construct our coarse $M_*$ bins in
Fig.~\ref{fig:money} to mitigate the effects of uncertainty in $M_*$.

In the upper-panel of Fig.~\ref{fig:money}, we show the fraction of
objects in the mass-matched bins that are determined to be mid-IR AGN
($f_{\rm AGN}$). In the $0.3<z<0.6$ and $0.6<z<0.9$ bins, these AGN
fractions are corrected using our derived $M_*$ completeness
functions. We find that the AGN fractions in our major merger and
all-merger samples are a factor $\sim2-7$ higher than those for
non-interacting galaxies. This result appears to be consistent across
the redshift range considered here, in that $f_{\rm AGN}$ is
systematically higher for interacting galaxies over non-interacting
galaxies, and this in observed in each of the redshift bins. We find
that this enhancement in $f_{\rm AGN}$ is significant at the
3.5--5.4$\sigma$ level for the sources with $M_* < 10^{11} M_{\odot}$,
dropping to 1.7--2.2$\sigma$ in the high-mass bin for each individual
redshift slice considered here. When combining the data across the
redshift bins, the significance of this result increases to
3.3--8.0$\sigma$ across all stellar masses.

To further test the robustness of this result to the misidentification
of AGN in our sample (i.e., normal star-forming galaxies with
unusually red mid-IR colors that could mimick an AGN-signature), we
used a stricter W1-W2 color cut of $>1.0$ to classify a source as an
AGN, and re-calculated the AGN fractions. While this cut greatly
affected the number of AGN being identified, particularly in the
lowest-redshift bin, we found that the increase in $f_{\rm AGN}$ for
interacting galaxies was still significant in six of the nine $z-M_*$
bins.

The fact that we observe an increase in $f_{\rm AGN}$ in each redshift
bins suggests this result is independent of our $M_*$ completeness
corrections. Indeed, we find that if we do not implement a
completeness correction, $f_{\rm AGN}$ remains enhanced by a similar
factor in merging/interacting systems. For interacting systems, we
find a marginal increase in $f_{\rm AGN}$ with $M_*$ that is more
pronounced with increasing redshift. However, we find that the
fractional difference of $f_{\rm AGN}$ for mergers and non-interacting
galaxies does not appear to be conditional on $M_*$.

Consistent with the merging/interacting galaxies, at $z>0.3$, there is
a marginal enhancement in $f_{\rm AGN}$ for the non-interacting
systems with the largest $M_*$ ($> 10^{11} M_{\odot}$) over the lower
mass non-interacting galaxies. However, we note that only 2 AGN are
identified in the highest mass bin for the isolated systems at
$z<0.3$; these poor source statistics would prevent us from
significantly identifying a similar rise at high masses, as observed
in the higher redshift bins. Although we observe a rise in AGN
fraction related to $M_*$ for isolated systems, crucially, these
measurements do not exceed the AGN fractions found for interacting
galaxies at the same $M_*$.

\subsection{Testing for observational bias and heterogeneity in our
  spec-$z$ sample}
\label{sec:obsbias}

As discussed in Section~\ref{sec:speczsample}, our parent galaxy
sample is constructed from a heterogeneous set of spectroscopic
redshift surveys. While the majority of the surveys targeted all
galaxies to a given brightness threshold and within a particular
region of the sky, spectroscopic redshifts may still not have been
measured for some objects. This can be due to observing difficulties,
signal to noise effects, lack of emission features etc., and hence are
individually incomplete at some level to all galaxies within the sky
region. Moreover, spectroscopic surveys such as SDSS-BOSS invoke
optical color cuts to pre-select galaxies in a given redshift range,
which results in complex selection/incompleteness effects. Thus, each
spectroscopic redshift survey has its own unique set of selection
biases, which become imprinted onto the main parent galaxy sample
considered throughout our analyses. Here we test whether the
spectroscopic surveys are biasing the AGN fractions measured in the
previous section and presented in Figure~\ref{fig:money}.

In Figure~\ref{fig:fig1} we showed that our parent sample is dominated
by objects drawn from 2--4 different spectroscopic surveys for each of
the three redshifts bins (i.e., $0.1<z<0.3$; $0.3<z<0.6$; $0.6<z<0.9$)
considered throughout. To test whether one of the spectroscopic
surveys excessively contributes to the measured AGN fractions in any
of the redshift bins presented in Figure~\ref{fig:money}, and hence
may be causing a bias in the AGN fraction at those redshifts, we
systematically removed all objects pertaining to one particular
redshift survey and recomputed the AGN fractions for that redshift
bin. For example, in the lowest redshift bin at $0.1<z<0.3$, it is
clear that our parent sample is mainly drawn from objects presented in
the SDSS-Legacy and GAMA surveys. Hence, we removed all galaxies
(irrespective of morphology) drawn from the SDSS-Legacy survey and
recomputed the AGN fractions at $0.1<z<0.3$ for the three
morphology/interaction categories for a single stellar mass
bin.\footnote{The significantly lower number of sources at a given
  redshift caused by removing a particular redshift survey were not
  sufficient to allow a statistically significant separation into
  multiple bins of stellar mass} Even after removing the SDSS-Legacy
survey objects, at $0.1<z<0.3$, we found fully consistent AGN
fractions with those presented in Fig.~\ref{fig:money}:
$f_{\rm AGN,Major-Merger}\sim 0.031$,
$f_{\rm AGN,All-Merger}\sim 0.029$ and
$f_{\rm AGN,Non-Interacting} \sim 0.0065$. We repeated this test at
$0.1<z<0.3$ by removing all objects drawn from the GAMA survey, and
again found a factor $\sim 3.5$ increase in the interacting AGN
fractions over the non-interacting sources. We continued this test in
each redshift bin for each redshift survey, and consistently found
that AGN are more prevalent in interacting galaxies and major mergers
than non-interacting systems. The only marginal bias we observed
during this test was with the exclusion of SDSS-BOSS galaxies, where
we found that the AGN fractions difference between non-interacting and
interacting galaxies increased from a factor $\sim 3$ to a factor
$\sim 6$ at $0.3<z<0.6$. This suggests a possible bias against the
targeting of merging galaxies and/or AGN in the SDSS-BOSS survey, and
hence, the $f_{\rm AGN}$ presented in Fig.~\ref{fig:money} at
$0.3<z<0.6$ may be marginally conservative, and the true difference
between mergers and non-interacting galaxies is likely to be larger.

\subsection{The most luminous AGN preferentially reside in merging
  galaxies}

\begin{figure}
  \begin{center}
    \includegraphics[width=\linewidth]{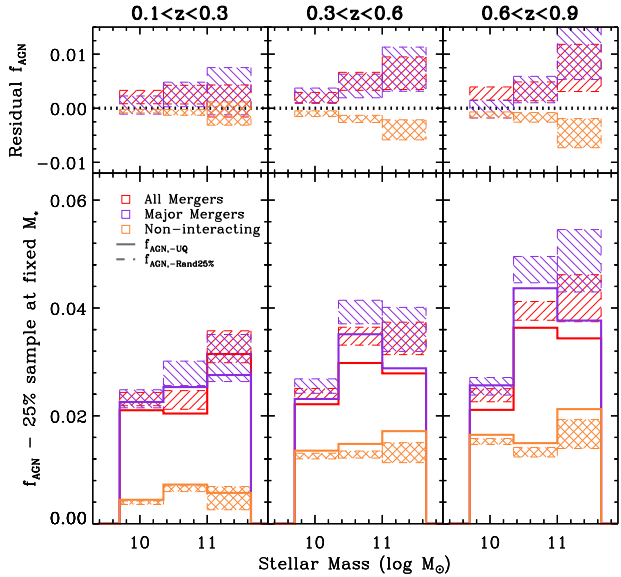} 
  \end{center}
  \caption{{\bf Lower Panels:} AGN fraction as a function of stellar
    mass after removal of the upper quartile ($f_{\rm AGN,-UQ}$) of
    sources with the highest $L_{\rm AGN}$ (solid histograms). Hashed
    regions provide the AGN fraction determined after randomly
    removing 25\% ($f_{\rm AGN,-Rand25}$) of the AGN in a given
    stellar mass bin, irrespective of $L_{\rm AGN}$. This fraction is
    recomputed 10,000 times using a jack-knife re-sampler, and the
    bounds provide the 90th percentile range of the samples. Color
    coding is the same as Fig~\ref{fig:money}. {\bf Upper Panels:}
    Residual AGN fraction between the $f_{\rm AGN,-UQ}$ and
    $f_{\rm AGN,-Rand25}$ (i.e.,
    $\Delta f_{\rm AGN} = f_{\rm AGN,-Rand25} - f_{\rm AGN,-UQ}$) as a
    function of stellar mass. In general, we find that the most
    luminous AGN systematically reside in the interacting/merging
    galaxies, hence their positive residual signatures. Conversely,
    fewer luminous AGN reside in non-interacting galaxies, as we show
    that $f_{\rm AGN,-UQ}$ is systematically larger than the expected
    null result value,
    $f_{\rm AGN,-Rand25}$.}\label{fig:fAGN_residuals}
\end{figure}

Several recent studies have identified a possible correlation between
AGN luminosity and galaxies undergoing mergers. These studies found
high merger fractions ($\sim 85\%$) in luminous
$L_{\rm AGN} > 10^{46}$~erg~s$^{-1}$ dust reddened quasars (e.g.,
\citealt{Urrutia:2008aa,Glikman:2012aa}), consequently leading to the
intriguing suggestion that merger fraction is dependent on AGN
bolometric luminosity \citep{Treister:2012aa}. This has been
substantiated by the identification of a strong positive trend of
increasing merger fraction spanning over 3 decades in $L_{\rm AGN}$,
which may persist beyond $z \sim 2$ (e.g.,
\citealt{Kocevski:2015aa,Del-Moro:2016aa,Fan:2016aa}). However, others
have found less convincing evidence for a connection between merger
fraction and the most luminous AGN activity, particularly in Type-1
quasars (e.g., \citealt{Villforth:2017aa}). Furthermore, at more
moderate luminosities, $L_{\rm AGN} < 10^{44}$~erg~s$^{-1}$, this
connection appears to be weaker, with merger fractions remaining
relatively constant (at $\sim 20\%$) with decreasing $L_{\rm AGN}$.
These seemingly contradictory results are consistent with a picture in
which mergers drive the high-Eddington growth of the most massive BHs
($L_{\rm AGN} \gtrsim 10^{44}$~erg~s$^{-1}$), but with
secular/internal processes becoming increasingly dominant at lower
Eddington ratios and/or lower BH masses.

We can test such a scenario using our interacting and non-interacting
galaxy samples by assessing whether the most bolometrically luminous
AGN are preferentially hosted in interacting galaxies over
non-interacting galaxies. If the most luminous AGN are biased
regarding the morphologies/interaction state of the host, then the AGN
fractions for interacting/non-interacting galaxies, presented in
Fig.~\ref{fig:money}, should have an additional dependency on
$L_{\rm AGN}$ at fixed stellar mass. However, and crucially, if indeed
the most luminous AGN are systematically more likely to reside in
merging galaxies, then $f_{\rm AGN, interacting}$ and
$f_{\rm AGN, non-interacting}$ would not have the $same$ dependency on
$L_{\rm AGN}$, at fixed stellar mass.

We test this hypothesis by measuring the change in $f_{\rm AGN}$
between those values presented in Fig.~\ref{fig:money}, and the
$f_{\rm AGN}$ calculated after removing the upper-quartile of the most
luminous AGN present in a given $M_*$ bin (hereafter,
$f_{\rm AGN,-UQ}$)\footnote{Our choice of selecting the upper 25\% of
  the AGN luminosities is somewhat arbitrary. However, this ensures
  that we remove the majority of AGN with
  $L_{\rm AGN} > 10^{45}$~erg~s$^{-1}$, which is consistent with the
  upturn in the merger fraction as a function of $L_{\rm AGN}$ (e.g.,
  \citealt{Fan:2016aa} and refs. therein).} For consistency, we use
the same bins of $M_*$ and $z$ as those presented in
Fig.~\ref{fig:money}. Similar to our analysis in
Section~\ref{sec:incidence}, by measuring changes in $f_{\rm AGN}$
between interacting and non-interacting galaxies, we naturally control
for BH accretion variability, which likely occurs on shorter
timescales than merger events.

AGN bolometric luminosities for our sample are computed by fitting
powerlaw slopes to the mid-IR photometry of the AGN, and using the
best-fit slope to predict the rest-frame $6 \mu m$ continuum
luminosity, which is shown to be a robust indicator of $L_{\rm AGN}$
(e.g., \citealt{Lutz:2004aa,Fiore:2009aa,Chen:2017ac}). The AGN
considered here cover a relatively wide range, with luminosities of
$L_{\rm AGN} \sim 3 \times 10^{43}$--$2 \times 10^{46}$~erg~s$^{-1}$.

In each $M_*$ bin, we find that the AGN fraction decreases
substantially for the major-merger and interacting galaxy samples by
factors of $\sim 0.7-2$ after removing the upper-quartile of the most
luminous AGN, i.e.
$\Delta f_{\rm AGN} = f_{\rm AGN} - f_{\rm AGN,-UQ} \sim 0.01 - 0.03$.
While the decrease in $f_{\rm AGN}$ for the non-interacting galaxies
was in some cases consistent with $\sim 0$. This provides tentative
evidence that the most luminous AGN do preferentially reside in
merging galaxies.

We can give these results a stronger statistical footing by simulating
the $\Delta f_{\rm AGN}$ had we not preferentially removed the upper
quartile of the most luminous AGN, but instead we had randomly removed
25\% of the AGN from a given $M_*$ bin (i.e., independent of
$L_{\rm AGN}$). This is achieved through 10,000 Jack-knife re-samplings
of $f_{\rm AGN}$ for each morphology/interaction sample after removing
a random set of 25\% of the AGN in each $M_*$ bin at each redshift
($f_{\rm AGN,-Rand25}$). In Fig.~\ref{fig:fAGN_residuals} we plot the
residual between $f_{\rm AGN,-Rand25}$ and $f_{\rm AGN,-UQ}$. The
hashed regions represent the 90th percentiles of the Jack-knife
samples. Across all three redshift ranges, we show a general trend of
positive $\Delta f_{\rm AGN}$ for the interacting galaxies (i.e.,
$f_{\rm AGN,-Rand25,mergers} > f_{\rm AGN,-UQ}$), and negative
residuals for the non-interacting galaxies (i.e.,
$f_{\rm AGN,-Rand25,non-interacting} < f_{\rm AGN,-UQ}$). Hence, we
have good statistical evidence at the $>90$\% level, that the most
luminous AGN systematically reside in the interacting galaxies at
fixed stellar mass.

In Fig.~\ref{fig:fAGN_residuals}, we further show that at $z>0.3$ the
$\Delta f_{\rm AGN}$ appears to diverge with increasing $M_*$,
suggesting that a larger fraction of the most luminous AGN reside in
interacting galaxies at higher $M_*$, with an additional marginal
preference for the most luminous AGN residing in major-mergers. At
$z<0.3$ we find consistent (at the 90th percentile) residual
$f_{\rm AGN}$ values between the interacting and non-interacting
galaxies in the highest $M_*$ bin, suggesting no statistically
significant preference for luminous AGN between the interaction
classifications. We note that this result may also be driven by the
small number of AGN in non-interacting galaxies at high $M_*$ in our
sample, preventing us from measuring a strong systematic
difference. However, overall we show that fewer luminous AGN reside in
non-interacting galaxies, with a strong preference for the most
rapidly growing BHs to be generally hosted in major mergers.

\section{Discussion}

\subsection{Evidence for stochastic BH growth during major-merger
  events}

Based on our population analysis (Section~\ref{sec:incidence}), we can
robustly conclude that, on average, those galaxies that are currently
undergoing or have recently undergone some form of a
merger/interaction are a factor $\sim2-7$ more likely to be rapidly
growing their central BHs than more isolated and non-interacting
star-forming galaxies. Furthermore, whilst we have not attempted to
address the question of whether all luminous AGN events must be
triggered by mergers (such a study would require a thorough
understanding of incompleteness effects), our results clearly indicate
a {\it systematic enhancement} of $f_{\rm AGN}$ in major mergers over
non-interacting galaxies and/or even minor-mergers. This could suggest
that significant BH growth phase(s) are linked specifically to a major
merger scenario.

Previous studies that have investigated merger--AGN connections have
produced mixed results
(e.g. \citealt{Gabor:2009aa,Cisternas:2011aa,Schawinski:2011aa,Kocevski:2012aa,Treister:2012aa,Ellison:2013aa,Villforth:2014aa,Kocevski:2015aa}),
often finding that the host galaxies of AGN are similar to those of
non-AGN (i.e., no strong enhancement in merger fractions of AGN over
non-AGN). These results are seemingly at odds with our population
analysis that shows, on average, AGN are markedly more likely to occur
during a major-merger than in an isolated galaxy. We suggest that
these apparent differences can be reconciled by considering the
relative time-scales of the luminous AGN activity, the dynamical time
of a major-merger, and the time spent as an isolated galaxy.

Motivated by galaxy merger simulations, here we outline a framework
that closely ties BH accretion rate variability to the dynamical time
of the galaxy merger. Given that AGN activity is known to vary on
timescales much shorter than galaxy processes (e.g.,
\citealt{Novak:2011aa,Gabor:2013aa}), and on average, BH accretion
rate is linked to available gas supply (e.g.,
\citealt{Hickox:2014aa}), we make the ansatz that the act of merging
further enhances AGN variability as galaxy merging strongly affects
the inflow of gas, which can serve to fuel the AGN. For example, on
first pericentric passage, gravitational torques may be sufficient to
induce a short period of rapid BH growth ($< 50$~Myrs). After first
passage, interaction signatures, such as tidal tails, may still be
evident as the galaxies move to maximum separation (lasting
$\sim 200-400$~Myrs).  However, internally, the galaxies may partially
relax, limiting fuel to the BH, and causing accretion to slow, and the
wide-separation merger may no longer be observationally identified as
an AGN. Indeed, the fraction of AGN is seen to fall dramatically at
large projected separations (e.g.,
\citealt{Ellison:2013aa,Satyapal:2014aa,Ricci:2017aa}). However, new
episodes of significant AGN activity may then be re-ignited on
subsequent passages until coalescence of the galaxies. In such a
scenario, the AGN activity would seemingly occur sporadically
throughout the merger, but with the overall AGN light curve being
strongly correlated with the merger's dynamical time.

Within our proposed framework, during the merger there may be multiple
periods of observable AGN activity, and seemingly non-AGN
activity. Including a close connection with the dynamical time forces
the non-AGN phases to last substantially longer during first and
second passage, but with the non-active phases become shorter-lived as
the galaxies begin to coalesce. In accordance with the results of
previous investigations, comparisons of the merger rates of AGN to
merger rates in non-AGN should produce similar fractions, as the AGN
is not always `active' during the entire merger event. As the merger
begins to reach coalescence, the probability to observe the AGN
increase dramatically as the AGN episodes occur more frequently, and
may result in the maximal growth phase of the BH, as predicted in
merger simulations. Such a scenario would simultaneously explain the
apparent increase in merger rate with AGN luminosity (e.g.,
\citealt{Urrutia:2008aa,Glikman:2012aa,Treister:2012aa}), and our
finding of the most luminous AGN preferentially residing within
merging systems.

Furthermore, in Figure~\ref{fig:money}, we show that $\lesssim 10$\%
of major-mergers contain luminous (obscured mid-IR) AGN, suggesting
that the total AGN duty-cycle over the course of the merger may only
be $\sim 50-100$~Myrs. Moreover, in Figure~\ref{fig:money}, we show
that luminous AGN activity still occurs during isolated, secular
evolution phases (i.e., a major-merger may not be an absolute
requirement for luminous AGN activity to occur). Taken together, these
two results suggest that merger-AGN studies primarily selected on the
basis of AGN activity would need to be large in number in order to
observe the relatively small $\sim 10$\% enhancement in mergers
relative to non-mergers in an AGN-selected sample (e.g.,
\citealt{Villforth:2014aa}). However, we show that when considering a
time-averaged look at the AGN duty cycle, the probability of luminous
AGN occurring in major-mergers is clearly enhanced over
non-interacting galaxies by at least a factor $\sim 3$.

Overall, our results suggest that, on average, mergers do trigger AGN
significantly more often than in secularly evolving galaxies above a
particular luminosity threshold
($L_{\rm AGN} \gtrsim 10^{44}$~erg~s$^{-1}$), but that the BH does not
necessarily need to be growing at a significant rate throughout the
entire merger phase. However, given that we have focused on the growth
of obscured AGN only, we have implictly not considered a possible
evolutionary scenario between Type 2 and Type 1, which may also be
linked with merger stage. This is beyond the scope of this
investigation, but may be possible with future wide-format
high-resolution imaging capable of identifying Type-1 AGN, while also
providing the ability to extract galaxy properties such as
morphology. Finally, at lower BH masses and/or lower AGN luminosities,
secular processes may be more important for driving BH growth, with
major-mergers becoming sub-dominant. Indeed, an enhancement in AGN
fraction of only a factor $\sim 0.5-2$ is seen in Seyfert-like
luminosity $z<0.1$ galaxy pairs in SDSS (e.g.,
\citealt{Ellison:2011aa}).

\section{Summary \& Conclusions}

In this paper, we have investigated the effect of the merging of
gas-rich galaxies on the growth of BHs out to $z \lesssim 1$. We have
used the exquisite imaging quality afforded to us by the HSC
instrument on the Subaru Telescope to identify merging and
non-interacting galaxies across the first 170~deg$^2$ of the HSC
survey. We used publicly available archival data within the HSC survey
regions to identify spectroscopically confirmed galaxies in the
redshift range $0.1 < z < 0.9$ (see Section~\ref{sec:speczsample}),
and performed SED fitting (see Section~\ref{sec:sedfitting}) to derive
their internal properties. We used photometry from the all-sky WISE
mid-IR survey to identify the galaxies in our sample containing
luminous AGN (see Section~\ref{sec:agnID}), and used the sensitive and
high spatial resolution HSC imaging to implement a Random Forest
machine learning algorithm to robustly identify large samples of
merging and non-interacting galaxies (see
Section~\ref{sec:randomforest}). We use our morphological
classifications in conjunction with the mid-IR AGN identifications to
place constraints on the average incidence of luminous AGN in merging
versus non-interacting galaxies. Our conclusions are the following:

\begin{enumerate}

\item Based on stellar mass matched samples of galaxies, BHs hosted in
  merging galaxies are a factor $\sim 2-7$ more likely to be rapidly
  growing than in non-interacting galaxies. This result is found to be
  consistent in three separate redshift bins ($0.1 < z < 0.3$;
  $0.3 < z < 0.6$; $0.6 < z < 0.9$), and is relatively independent of
  stellar mass.

\item Our parent sample of galaxies is drawn heterogeneously based on
  spectroscopic redshift confirmations from a variety of dedicated
  surveys. We investigated the likelihood of our AGN fractions being
  driven by the source selection induced by any one of these
  spectroscopic surveys by systematically removing each individual
  redshift survey from our parent sample and recomputing the AGN
  fractions as a function of morphology. We determined that our
  results are not driven by spectroscopic selection, finding fully
  consistent AGN fractions throughout.

\item Several previous studies have suggested a strong link between
  galaxy merging and the most luminous AGN. We tested this result by
  assessing whether the most bolometrically luminous AGN are
  systematically hosted in merging galaxies over non-interacting
  systems. At any given stellar mass bin, we found that the
  upper-quartile of the most luminous AGN preferentially reside in
  merging galaxies over non-interacting galaxies. We use these results
  to suggest that a major merger between two galaxies is sufficient to
  induce a flow of cool gas towards the central BH in one or both
  galaxies, and this is systematically more likely to trigger a
  significant AGN event that in an isolated galaxy alone.

\item To place our findings into the wider context of AGN--galaxy
  co-evolution, and reconcile our conclusions with seemingly
  contradictory results within the recent literature, we outline a
  coherent framework that closely ties the variable AGN light curve to
  the dynamical time of the merger event. Our proposed framework
  requires that AGN accretion undergoes several distinct peaks in
  luminosity over the lifetime of the merger, with BH fueling linked
  to the close passage and interaction of the merging galaxies. The
  substantial time spent at wide pair separations, when the BH is not
  growing at an appreciable rate, serves to explain previous findings
  that highlight similarities between the fractions of AGN and non-AGN
  in merger states.

\end{enumerate}

\noindent Overall, our morphological investigation of $0.1<z<0.9$
galaxies identified in the first 170~deg$^2$ of the HSC survey,
provides conclusive evidence that luminous AGN are systematically more
likely (by at least a factor $\gtrsim 3$) to occur in major-mergers
when compared to non-interacting galaxies. Our results suggest that,
on average, mergers do trigger AGN significantly more often than in
secularly evolving galaxies. However, the BH need not be growing at an
appreciable rate throughout the entire merger phase.

\begin{ack}
  We thank the anonymous referee for their considered report, which
  allowed us to clarify and improve several aspects of this
  manuscript.  ADG and JEG gratefully acknowledge support from the
  National Science Foundation under Grant Number AST-1613744. YM was
  supported by JSPS KAKENHI Grant No. JP17H04830.  The authors thank
  Lisa Kewley, Robert Lupton, Jim Bosch and Bob Armstrong for
  enlightening conversations. This publication makes use of data
  products from the Wide-field Infrared Survey Explorer, which is a
  joint project of the University of California, Los Angeles, and the
  Jet Propulsion Laboratory/California Institute of Technology, funded
  by the National Aeronautics and Space Administration. The Hyper
  Suprime-Cam (HSC) collaboration includes the astronomical
  communities of Japan and Taiwan, and Princeton University. The HSC
  instrumentation and software were developed by the National
  Astronomical Observatory of Japan (NAOJ), the Kavli Institute for
  the Physics and Mathematics of the Universe (Kavli IPMU), the
  University of Tokyo, the High Energy Accelerator Research
  Organization (KEK), the Academia Sinica Institute for Astronomy and
  Astrophysics in Taiwan (ASIAA), and Princeton University. Funding
  was contributed by the FIRST program from Japanese Cabinet Office,
  the Ministry of Education, Culture, Sports, Science and Technology
  (MEXT), the Japan Society for the Promotion of Science (JSPS), Japan
  Science and Technology Agency (JST), the Toray Science Foundation,
  NAOJ, Kavli IPMU, KEK, ASIAA, and Princeton University.

\end{ack}

\footnotesize{
\bibliographystyle{apj}
\bibliography{submitted.bbl}
}

\end{document}